\documentclass[11pt]{article}

\usepackage[top=1in,bottom=1in,left=1in,right=1in]{geometry}
\usepackage{natbib}
\usepackage[unicode=true,bookmarks=true,bookmarksnumbered=false,bookmarksopen=false,breaklinks=false,pdfborder={0 0 1},backref=false,colorlinks=true]{hyperref}
\hypersetup{citecolor=blue}
\usepackage{url}

\usepackage{amsmath}
\usepackage{mathrsfs}
\usepackage{amssymb}
\usepackage{amsthm}
\usepackage[normalem]{ulem}

\newtheorem{theorem}{Theorem}
\newtheorem{lemma}{Lemma}

\usepackage{float}
\usepackage{rotfloat}

\floatstyle{ruled}
\newfloat{algorithm}{tpb}{loa}
\providecommand{\algorithmname}{Algorithm}
\floatname{algorithm}{\protect\algorithmname}

\usepackage{algpseudocode,algorithm}
\algnewcommand\algorithmicinput{\textbf{Input}:}
\algnewcommand\algorithmicoutput{\textbf{Output}:}
\algnewcommand\INPUT{\item[\algorithmicinput]}
\algnewcommand\OUTPUT{\item[\algorithmicoutput]}

\usepackage{graphicx}
\usepackage[caption=false,labelformat=simple]{subfig}

\usepackage{array}
\newcolumntype{L}[1]{>{\raggedright\let\newline\\\arraybackslash\hspace{0pt}}m{#1}}
\newcolumntype{C}[1]{>{\centering\let\newline\\\arraybackslash\hspace{0pt}}m{#1}}
\newcolumntype{R}[1]{>{\raggedleft\let\newline\\\arraybackslash\hspace{0pt}}m{#1}}

\usepackage{rotating}
\usepackage{multirow}

\usepackage{tikz}
\usetikzlibrary{shapes,arrows,backgrounds,calc,positioning,fit,petri,plotmarks}
\usetikzlibrary{arrows}

\usepackage{enumerate}
\usepackage{paralist}

\newcommand*{\affaddr}[1]{#1} % No op here. Customize it for different styles.
\newcommand*{\affmark}[1][*]{\textsuperscript{#1}}

\global\long\def\bx{\mathbf{x}}
\global\long\def\bY{\mathbf{Y}}
\global\long\def\by{\mathbf{y}}
\global\long\def\bZ{\mathbf{Z}}
\global\long\def\bz{\mathbf{z}}

\global\long\def\bA{\mathbf{A}}

\global\long\def\bH{\mathbf{H}}

\global\long\def\bw{\mathbf{w}}

\global\long\def\bQ{\mathbf{Q}}

\global\long\def\bS{\mathbf{S}}

\global\long\def\bgamma{\boldsymbol{\gamma}}
\global\long\def\bbeta{\boldsymbol{\beta}}
\global\long\def\bpi{\boldsymbol{\pi}}
\global\long\def\bGamma{\boldsymbol{\Gamma}}

\usepackage{xr}
\makeatletter
\newcommand*{\addFileDependency}[1]{% argument=file name and extension
  \typeout{(#1)}
  \@addtofilelist{#1}
  \IfFileExists{#1}{}{\typeout{No file #1.}}
}
\makeatother
 
\newcommand*{\myexternaldocument}[1]{%
    \externaldocument{#1}%
    \addFileDependency{#1.tex}%
    \addFileDependency{#1.aux}%
}

\myexternaldocument{}
\title{Longitudinal regression of covariance matrix outcomes}

\author{%
    Yi Zhao\affmark[1], Brian S. Caffo\affmark[2], Xi Luo\affmark[3], and
    for the Alzheimer's Disease Neuroimaging Initiative\footnote{Data used in preparation of this article were obtained from the Alzheimer's Disease Neuroimaging Initiative (ADNI) database (\url{adni.loni.usc.edu}). As such, the investigators within the ADNI contributed to the design and implementation of ADNI and/or provided data but did not participate in analysis or writing of this report. A complete list of ADNI investigators can be found at: \url{http://adni.loni.usc.edu/wp-content/uploads/how_to_apply/ADNI_Acknowledgement_List.pdf}} \\
    \affaddr{\affmark[1]Department of Biostatistics and Health Data Science, Indiana University School of Medicine} \\
    \affaddr{\affmark[2]Department of Biostatistics, Johns Hopkins Bloomberg School of Public Health} \\
    \affaddr{\affmark[3]Department of Biostatistics and Data Science,\\  The University of Texas 
Health Science Center at Houston} \\
}

\date{}

\providecommand{\keywords}[1]
{
  {\small	
  \textbf{Keywords:} #1 Covariance regression; Hierarchical likelihood; Multilevel model; Shrinkage estimator}
}
\begin{document}
%%%%%%%%%%%%%%%%%%%%%%%%%%%%%%%%%%%%%%%%%%%%%%%%%%%%%%%%%%

\maketitle

\thispagestyle{empty}

\begin{abstract}
In this study, a longitudinal regression model for covariance matrix outcomes is introduced. The proposal considers a multilevel generalized linear model for regressing covariance matrices on (time-varying) predictors. This model simultaneously identifies covariate associated components from covariance matrices, estimates regression coefficients, and estimates the within-subject variation in the covariance matrices. Optimal estimators are proposed for both low-dimensional and high-dimensional cases by maximizing the (approximated) hierarchical likelihood function and are proved to be asymptotically consistent, where the proposed estimator is the most efficient under the low-dimensional case and achieves the uniformly minimum quadratic loss among all linear combinations of the identity matrix and the sample covariance matrix under the high-dimensional case. Through extensive simulation studies, the proposed approach achieves good performance in identifying the covariate related components and estimating the model parameters. Applying to a longitudinal resting-state fMRI dataset from the Alzheimer's Disease Neuroimaging Initiative (ADNI), the proposed approach identifies brain networks that demonstrate the difference between males and females at different disease stages. The findings are in line with existing knowledge of AD and the method improves the statistical power over the analysis of cross-sectional data.
\end{abstract}

%%%%%%%%%%%%%%%%%%%%%%%%%%%%%%%%%%%%%%%%%%%%%%%%%%%%%%%%%%
%========================================================%

%========================================================%

%%%%%%%%%%%%%%%%%%%%%%%%%%%%%%%%%%%%%%%%%%%%%%%%%%%%%%%%%%
%========================================================%
\clearpage
\setcounter{page}{1}

%%%%%%%%%%%%%%%%%%%%%%%%%%%%%%%%%%%%%%%%%%%%%%%%%%%%%%%%%%
% Introduction
%%%%%%%%%%%%%%%%%%%%%%%%%%%%%%%%%%%%%%%%%%%%%%%%%%%%%%%%%%
\section{Introduction}
\label{sec:intro}

This manuscript introduces a longitudinal principal regression model for multiple covariance matrix outcomes. For cross-sectional data, a generalized regression model for the covariance matrix outcomes with a logarithmic link function has been proposed:
\begin{equation}\label{eq:model_crosssection}
  \log(\bgamma^\top\Sigma_{i}\bgamma)=\bx_{i}^\top\bbeta,
\end{equation}
where $\Sigma_{i}\in\mathbb{R}^{p\times p}$ is an assumed true covariance matrix of subject $i$ and $\bx_{i}\in\mathbb{R}^{q}$ is a vector of covariates, for $i=1,\dots,n$. The parameter vector, $\bgamma\in\mathbb{R}^{p}$, is a linear projection and $\bbeta\in\mathbb{R}^{q}$ is the model coefficient, both to be estimated. Model~\eqref{eq:model_crosssection} was first introduced in \citet{zhao2021covariate} for the small $p$ setting for characterizing brain functional connectivity (represented by the covariance matrix of the data) with individual/population characteristics in resting-state functional magnetic resonance imaging (fMRI) studies. It has the advantage of directly identifying the networks that are associated with the covariates and, at the same time, offering relatively high flexibility in the model formulation.
Later, it was extended to handle higher dimensional cases either through a dimension-reduction step~\citep{zhao2021whole} or by introducing a shrinkage estimator of the covariance matrices~\citep{zhao2021principal}.
These methods were developed for cross-sectional studies, and thus, were not optimal for analyzing and contrasting longitudinal imaging studies.
% increasingly popular longitudinal imaging studies.

In neuroimaging studies, understanding functional changes in and between healthy and pathological brains is an essential topic. With repeated measurements, longitudinal analysis enables the assessment of within-individual changes as well as the articulation of systematic differences among individuals. One representative example is neurodegenerative studies. With demographic shifts in aging, Alzheimer's disease (AD) and related dementias are a major public health challenge. Understanding disease pathology, identifying biological markers, and suggesting early diagnosis and intervention strategies are critically important. The landmark Alzheimer's Disease Neuroimaging Initiative (ADNI) study aims to identify biomarkers for the early detection and tracking of AD to assist in the development of prevention and intervention strategies. In the study, longitudinal data were collected from various measurement domains, including: clinical, genetic, imaging, and biospecimen, after obtaining informed consent. Motivated by this longitudinal resting-state fMRI dataset, we propose to extend Model~\eqref{eq:model_crosssection} to appropriately integrate information collected at multiple visits to increase the statistical power of identifying covariate-related effects.

A classic way of analyzing longitudinal neuroimaging data is to extract voxel/regional data first and subsequently fit them with a univariate longitudinal analysis model, such as a hierarchical or mixed effects model or generalized estimating equations, one at a time, and then finish with a multiple testing correction procedure~\cite[see a review by][]{madhyastha2018current}. In fMRI studies, the study interest lies more in the exploration of the interactions between voxels/regions or the network-level properties~\citep{li2009lstgee,dai2017predicting}. Univariate approaches disregard network information and structured constraints in the data. For resting-state fMRI data, the covariance matrix of the signals is generally used to reveal the coactivation between units, so-called functional connectivity~\citep{friston2011functional}. Running longitudinal models on an individual element of the matrix ignores the positive definiteness resulting in a large number of hypothesis testing, which is deficient in statistical power, and prevents the ability to predict a valid functional connectivity matrix. Multivariate approaches, including principal component analysis (PCA) and independent component analysis (ICA), are generally applied for dimension reduction. However, investigations of longitudinal effects on brain networks are rare. Recently, a hierarchical ICA model was proposed to analyze longitudinal fMRI data, which enables the study of time-dependent effects on the IC decomposition~\citep{wang2019hierarchical}. Graph theory is a technique widely used in resting-state fMRI studies to reveal the topological architecture of brain networks. With longitudinal data, it offers a way of studying the temporal variations in the topological structures~\citep{madhyastha2018current}. However, the method can be sensitive to the definition of the graphs, where high heterogeneity may exist due to various reasons, such as the choice of brain parcellation and the statistical variation in graph estimation~\citep{farahani2019application}.
For the purpose of predicting a behavioral outcome, machine learning techniques, such as support vector machines, random forests, and neural networks, together with cross-validation are widely implemented. However, these approaches usually ignore the temporal dependency in the repeated measures. Thus, they cannot be used to reveal the with-in subject variation and track longitudinal changes~\citep{telzer2018methodological}.

In this study, we focus on identifying covariate related brain networks in a longitudinal setting. Thus, it is assumed that the linear projection, $\bgamma$, in~\eqref{eq:model_crosssection} is a constant over time. A multilevel model is proposed to capture the within-subject variation in the covariance matrix, where Model~\eqref{eq:model_crosssection} is adopted as the base-level model. Under normality assumptions, a likelihood-based approach is introduced to estimate the model parameters. For the case with high-dimensional data, by generalizing the proposal in \citet{zhao2021principal}, a linear shrinkage estimator of the covariance matrices is introduced, where the shrinkage parameter is assumed to be common across subjects and visits. By doing so, the estimator achieves the optimal property with the uniformly minimum quadratic loss asymptotically among all linear combinations of the identity matrix and the sample covariance matrix.

The rest of the paper is organized as the following. Section~\ref{sec:model} introduces the longitudinal regression model for covariance matrix outcomes. The estimation method is proposed and the asymptotic properties are studied. In Section~\ref{sec:sim}, the performance of the proposed approach is demonstrated through simulation studies. In Section~\ref{sec:adni}, the model is applied to a longitudinal resting-state fMRI data set collected by the Alzheimer's Disease Neuroimaging Initiative (ADNI). Section~\ref{sec:discussion} summarizes this manuscript with discussions. The technical proofs and additional analytical results are collected in the supplementary materials.
%%%%%%%%%%%%%%%%%%%%%%%%%%%%%%%%%%%%%%%%%%%%%%%%%%%%%%%%%%

%%%%%%%%%%%%%%%%%%%%%%%%%%%%%%%%%%%%%%%%%%%%%%%%%%%%%%%%%%
% Model
%%%%%%%%%%%%%%%%%%%%%%%%%%%%%%%%%%%%%%%%%%%%%%%%%%%%%%%%%%
\section{Model and Methods}
\label{sec:model}

Let $\by_{ivt}\in\mathbb{R}^{p}$ denote the $t$th $p$-dimensional outcome acquired from subject $i$ at visit $v$, for $t=1,\dots,T_{iv}$, $v=1,\dots,V_{i}$, and $i=1,\dots,n$, where $T_{iv}$ is the total number of observations, $V_{i}$ is the number of visits of subject $i$, and $n$ is the number of subjects. The outcome, $\by_{ivt}$, is assumed to follow a multivariate normal distribution with mean zero and covariance matrix $\Sigma_{iv}$. Without loss of generality, the distribution mean is assumed to be zero, as the study focus is to model the heterogeneity in the covariance matrix. In practice, this assumption can be satisfied by centering the data to zero. Denote $\bx_{iv}\in\mathbb{R}^{q}$ as the $q$-dimensional covariates of interest, where the covariates can vary by time. It is assumed that there exists a linear projection $\bgamma\in\mathbb{R}^{p}$ such that the following model holds:
% \begin{eqnarray}\label{eq:model}
%   \log(\bgamma^\top\Sigma_{iv}\bgamma) &=& \beta_{0i}+\bx_{iv}^\top\bbeta_{1}, \nonumber \\
%   % \beta_{0i} &\sim& \mathcal{N}(\beta_{0},\sigma^{2}),
%   \beta_{0i} &=& \beta_{0}+\epsilon_{i},
% \end{eqnarray}
\begin{equation}\label{eq:model}
  \log(\bgamma^\top\Sigma_{iv}\bgamma)=\beta_{0}+\bx_{iv}^\top\bbeta_{1}+u_{i},
\end{equation}
where $\beta_{0}\in\mathbb{R}$ is the intercept and $\bbeta_{1}\in\mathbb{R}^{q}$ is a fixed effect parameter.
% $\beta_{0i}$'s are assumed to follow a linear model centered at $\beta_{0}$ with error $\epsilon_{i}$
The residual, $u_{i}$, is normally distributed with mean zero and variance $\sigma^{2}$, for $i=1,\dots,n$. Denote $\beta_{0i}=\beta_{0}+u_{i}$. It is the random intercept in a (generalized) mixed effects model.
In this study, we assume that the linear projection, $\bgamma$, is constant over visits focusing on the investigation of the variation between subjects and/or visits within a specific subnetwork.

Here, we do not impose any structural assumption on the covariance matrices. Rather, we only assume that there exists at least one common linear projection that satisfies~\eqref{eq:model}.  For the case of high dimensionality, in order to yield a consistent estimate of the covariance matrix, structural assumptions, such as bandable covariance matrices, sparse covariance matrices, spiked covariance matrices, covariances with a tensor product structure, and latent graphical models, are generally imposed in many regularization-based methods~\citep{cai2016structured}. In the next section, we will introduce a shrinkage estimator of the covariance matrices, which does not require any structural assumption on the covariance matrices. In addition, the estimator is guaranteed to be positive definite, preserves the eigenstructure of the covariance matrices, and is easy to compute based on a simple and explicit formula.

%%%%%%%%%%%%%%%%%%%%%%%%%%%%%%%%%%%%%%%%%%%%%%%%%%%%%%%%%%

%%%%%%%%%%%%%%%%%%%%%%%%%%%%%%%%%%%%%%%%%%%%%%%%%%%%%%%%%%
% Methods
%%%%%%%%%%%%%%%%%%%%%%%%%%%%%%%%%%%%%%%%%%%%%%%%%%%%%%%%%%
\subsection{Methods}
\label{sub:method}

Under Model~\eqref{eq:model}, it is proposed to estimate the parameters using an approximation of the negative hierarchical-likelihood function:
\begin{equation}\label{eq:hll}
  \ell = \sum_{i=1}^{n}\sum_{v=1}^{V_{i}}\frac{T_{iv}}{2}\left\{(\beta_{0i}+\bx_{iv}^\top\bbeta_{1})+(\bgamma^\top\hat{\Sigma}_{iv}\bgamma)\exp(-\beta_{0i}-\bx_{iv}^\top\bbeta_{1})\right\}+\sum_{i=1}^{n}\left\{\frac{1}{2}\log\sigma^{2}+\frac{(\beta_{0i}-\beta_{0})^{2}}{2\sigma^{2}}\right\},
\end{equation}
where $\hat{\Sigma}_{iv}$ is an estimate of the covariance matrix $\Sigma_{iv}$. Replacing $\hat{\Sigma}_{iv}$ with the sample covariance matrix, denoted as $\bS_{iv}$, the first part in~\eqref{eq:hll} is the conditional likelihood given $\beta_{0i}$, and the second part is the likelihood function of $\beta_{0i}$. We consider using this approximate hierarchical likelihood function rather than marginalizing over $\beta_{0i}$, as the explicit solution is analytically and computationally inconvenient. In addition, maximizing the hierarchical likelihood function is asymptotically equivalent to maximizing the standard likelihood function~\citep{lee1996hierarchical}.

To avoid degeneration in $\bgamma$, it is proposed to optimize the following problem:
\begin{eqnarray}\label{eq:obj_func}
  \text{minimize} && \ell \nonumber \\
  \text{such that} && \bgamma^\top\bH\bgamma=1,
\end{eqnarray}
where $\bH$ is a positive-definite matrix. We set $\bH=\sum_{i=1}^{n}\sum_{v=1}^{V_{i}}T_{iv}\hat{\Sigma}_{iv}/\sum_{i=1}^{n}\sum_{v=1}^{V_{i}}T_{iv}$.

For the case of high-dimensional data, $\max_{i,v}T_{iv}\ll p$ with $p$ increasing to infinity, the sample covariance matrices are rank-deficient. The estimate of the eigenvalues and eigenvectors can be largely biased~\citep{johnstone2009sparse} and optimizing~\eqref{eq:obj_func} is numerically unstable. Generalizing the shrinkage estimator proposed in \citet{zhao2021principal} to a longitudinal setting, the solution to the following optimization problem is considered as an estimate of the covariance matrices.
\begin{eqnarray}\label{eq:opt_covariance}
  \underset{(\mu,\rho)}{\text{minimize}} && \frac{1}{n}\sum_{i=1}^{n}\frac{1}{V_{i}}\sum_{v=1}^{V_{i}}\mathbb{E}\left\{\bgamma^\top\Sigma_{iv}^{*}\bgamma-\exp(\beta_{0i}+\bx_{iv}^\top\bbeta_{1})\right\}^{2}, \nonumber \\
  \text{such that} && \Sigma_{iv}^{*}=\rho\mu\boldsymbol{\mathrm{I}}+(1-\rho)\bS_{iv}.
\end{eqnarray}
Here, it is assumed that the shrinkage parameters, $\rho$ and $\mu$, are constant over visits and subjects.
One can also assume constant shrinkage parameters over visits for each subject. As demonstrated in Theorem~\ref{appendix:thm:asmp_optimality} in Appendix Section~\ref{appendix:sec:asmp_lp}, the empirical solution to~\eqref{eq:opt_covariance} gives the optimal estimator of the covariance matrix which yields the uniformly minimum quadratic loss asymptotically among all linear combinations of the identity matrix and the sample covariance matrix.

Denote $\bw_{iv}=(1, \bx_{iv}^\top)^\top\in\mathbb{R}^{q+1}$ and $\bbeta_{i}=(\beta_{0i},\bbeta_{1}^\top)^\top\in\mathbb{R}^{q+1}$, and, $\beta_{0i}+\bx_{iv}^\top\bbeta_{1}=\bw_{iv}^\top\bbeta_{i}$. The following theorem gives the solution to the optimization problem.
\begin{theorem}\label{thm:Sigma_solution}
  The solution to optimization problem~\eqref{eq:opt_covariance} is
    \begin{equation}
      \Sigma_{iv}^{*}=\frac{\psi^{2}}{\delta^{2}}\mu\boldsymbol{\mathrm{I}}+\frac{\phi^{2}}{\delta^{2}}\bS_{iv}, \quad \text{for } v=1,\dots,V_{i}, ~i=1,\dots,n,
    \end{equation}
  and the minimum value is
    \[
      \min~\frac{1}{n}\sum_{i=1}^{n}\frac{1}{V_{i}}\sum_{v=1}^{V_{i}}\mathbb{E}\left\{\bgamma^\top\Sigma_{iv}^{*}\bgamma-\exp(\beta_{0i}+\bx_{iv}^\top\bbeta_{1})\right\}^{2}=\frac{\psi^{2}\phi^{2}}{\delta^{2}},
    \]
  where
    \[
      \mu=\frac{1}{n(\bgamma^\top\bgamma)}\sum_{i=1}^{n}\frac{1}{V_{i}}\sum_{v=1}^{V_{i}}\exp(\bw_{iv}^\top\bbeta_{i}), ~\phi^{2}=\frac{1}{n}\sum_{i=1}^{n}\frac{1}{V_{i}}\sum_{v=1}^{V_{i}}\phi_{iv}^{2}, ~\psi^{2}=\frac{1}{n}\sum_{i=1}^{n}\frac{1}{V_{i}}\sum_{v=1}^{V_{i}}\psi_{iv}^{2}, ~\delta^{2}=\frac{1}{n}\sum_{i=1}^{n}\frac{1}{V_{i}}\sum_{v=1}^{V_{i}}\delta_{iv}^{2},
    \]
    \[
      \phi_{iv}^{2}=\left\{\mu(\bgamma^\top\bgamma)-\exp(\bw_{iv}^\top\bbeta_{i})\right\}^{2},~\psi_{iv}^{2}=\mathbb{E}\left\{\bgamma^\top\bS_{iv}\bgamma-\exp(\bw_{iv}^\top\bbeta)\right\}^{2}, ~\delta_{iv}^{2}=\mathbb{E}\left\{\bgamma^\top\bS_{iv}\bgamma-\mu(\bgamma^\top\bgamma)\right\}^{2}.
    \]
  For $\forall~v\in\{1,\dots,V_{i}\}$ and $\forall~i\in\{1,\dots,n\}$, $\delta_{iv}^{2}=\phi_{iv}^{2}+\psi_{iv}^{2}$, and thus, $\delta^{2}=\phi^{2}+\psi^{2}$.
\end{theorem}
The shrinkage parameters in Theorem~\ref{thm:Sigma_solution} are expectations. In practice, the following sample counterpart, $\bS_{iv}^{*}$, is used to replace $\hat{\Sigma}_{iv}$ in~\eqref{eq:hll}:
\begin{equation}
  \bS_{iv}^{*}=\frac{\hat{\psi}^{2}}{\hat{\delta}^{2}}\mu\boldsymbol{\mathrm{I}}+\frac{\hat{\phi}^{2}}{\hat{\delta}^{2}}\bS_{iv}, \quad \text{for } v=1,\dots,V_{i}, ~ i=1,\dots,n,
\end{equation}
where
\[
  \hat{\delta}^{2}=\frac{1}{n}\sum_{i=1}^{n}\frac{1}{V_{i}}\sum_{v=1}^{V_{i}}\hat{\delta}_{iv}^{2}, ~\hat{\psi}^{2}=\frac{1}{n}\sum_{i=1}^{n}\frac{1}{V_{i}}\sum_{v=1}^{V_{i}}\min(\hat{\psi}_{iv}^{2},\hat{\delta}_{iv}^{2}), ~\hat{\phi}^{2}=\frac{1}{n}\sum_{i=1}^{n}\frac{1}{V_{i}}\sum_{v=1}^{V_{i}}(\hat{\delta}_{iv}^{2}-\hat{\psi}_{iv}^{2}),
\]
\[
  \hat{\delta}_{iv}^{2}=\left\{\bgamma^\top\bS_{iv}\bgamma-\mu(\bgamma^\top\bgamma)\right\}^{2}, ~\hat{\psi}_{iv}^{2}=\frac{1}{T_{iv}}\left\{\bgamma^\top\bS_{iv}\bgamma-\exp(\bw_{iv}^\top\bbeta_{i})\right\}^{2}, ~\hat{\phi}_{iv}^{2}=\hat{\delta}_{iv}^{2}-\hat{\psi}_{iv}^{2}.
\]
%==========================================
\subsection{Algorithm}
\label{sub:algorithm}

When $p<\min_{i,v}T_{iv}$, a natural choice of $\hat{\Sigma}_{iv}$ is the sample covariance matrix. However, in \citet{zhao2021principal}, through both simulation study and theoretical analysis, the shrinkage estimator achieves superior performance in estimating the covariance matrix with lower quadratic loss. Thus, we continue to use the shrinkage estimator when extending to longitudinal data. 
We consider a block coordinate descent algorithm to solve for the solutions. To avoid converging to a local minimum, we suggest to randomly choose a series of initial values and take the estimate with the lowest objective value of~\eqref{eq:hll}.
Algorithm~\ref{alg:covreg} summarizes the estimation procedure and Appendix Section~\ref{appendix:sec:alg_covreg} gives the details.
In the algorithm, Step 3 updates the shrinkage estimate of the covariance matrices. In Step 4, $\beta_{0i}$ ($i=1,\dots,n$) and $\bbeta_{1}$ are updated following the Newton-Raphson method. For the hyperparameters, $\beta_{0}$ and $\sigma^{2}$, they are updated by minimizing the negative hierarchical-likelihood function~\eqref{eq:hll} as
\[
  \beta_{0}^{(s+1)}=\frac{1}{n}\sum_{i=1}^{n}\beta_{0i}^{(s+1)}, \quad \sigma^{2(s+1)}=\frac{1}{n}\sum_{i=1}^{n}\left(\beta_{0i}^{(s+1)}-\beta_{0}^{(s+1)}\right)^{2}.
\]
For $\bgamma$, the update is the solution to the following optimization problem
\begin{eqnarray*}
  \text{minimize} && \bgamma^\top\left\{\sum_{i=1}^{n}\sum_{v=1}^{V_{i}}\frac{T_{iv}}{2}\exp\left(-\beta_{0i}^{(s+1)-\bx_{iv}^\top\bbeta_{1}^{(s+1)}}\right)\bS_{iv}^{*(s+1)}\right\}\bgamma, \\
  \text{such that} && \bgamma^\top\bH\bgamma=1,
\end{eqnarray*}
where the solution is provided in Algorithm 1 in \citet{zhao2021covariate}.
\begin{algorithm}
  \caption{\label{alg:covreg}The optimization algorithm for problems~\eqref{eq:obj_func} and~\eqref{eq:opt_covariance}.}
  \begin{algorithmic}[1]
    \INPUT $\{(\by_{iv1},\dots,\by_{ivT_{iv}}),\bx_{iv}~|~v=1,\dots,V_{i},~i=1,\dots,n\}$

    \State \textbf{initialization}: $(\bgamma^{(0)},\beta_{0i}^{(0)},\bbeta_{1}^{(0)},\beta_{0}^{(0)},\sigma^{2(0)})$

    \Repeat \; for iteration $s=0,1,2,\dots$

    \State  \; for $v=1,\dots,V_{i}$ and $i=1,\dots,n$, update
      \[
        \bS_{iv}^{*(s+1)}=\frac{\hat{\psi}^{2(s)}}{\hat{\delta}^{2(s)}}\mu^{(s)}\boldsymbol{\mathrm{I}}+\frac{\hat{\phi}^{2(s)}}{\hat{\delta}^{2(s)}}\bS_{iv},
      \]
      \; \; \; \; where $(\hat{\psi}^{2},\hat{\phi}^{2},\hat{\delta}^{2},\mu)$ are set to the value with $\bgamma=\bgamma^{(s)}$, $\beta_{0i}=\beta_{0i}^{(s)}$, and $\bbeta_{1}=\bbeta_{1}^{(s)}$,

    \State \; update $\bgamma$, $\beta_{0i}$, $\bbeta_{1}$, $\beta_{0}$, and $\sigma^{2}$ by solving~\eqref{eq:obj_func} with $\hat{\Sigma}_{iv}=\bS_{iv}^{*(s+1)}$, denoted as $\bgamma^{(s+1)}$, $\beta_{0i}^{(s+1)}$, $~~~~~~~~\bbeta_{1}^{(s+1)}$, $\beta_{0}^{(s+1)}$, and $\sigma^{2(s+1)}$, respectively,

    \Until{the objective function in~\eqref{eq:opt_covariance} converges;}

    \State consider a random series of initializations, repeat Steps 1--5, and choose the results with the minimum objective value.

    \OUTPUT $(\hat{\bgamma},\hat{\bbeta}_{1},\hat{\beta}_{0},\hat{\sigma}^{2})$
  \end{algorithmic}
\end{algorithm}

To determine the number of components, we generalize the metric of average deviation from diagonality introduced in \citet{zhao2021covariate} to the longitudinal setting. Let $\hat{\Gamma}^{(k)}\in\mathbb{R}^{p\times k}$ denote the first $k$ identified components, the metric is defined as
\begin{equation}
  \mathrm{DfD}(\hat{\Gamma}^{(k)})=\prod_{i=1}^{n}\prod_{v=1}^{V_{i}}\left[\frac{\det\left\{\mathrm{diag}(\hat{\Gamma}^{(k)\top}\hat{\Sigma}_{iv}\hat{\Gamma}^{(k)})\right\}}{\det(\hat{\Gamma}^{(k)\top}\hat{\Sigma}_{iv}\hat{\Gamma}^{(k)})}\right]^{T_{iv}/\sum_{i}\sum_{v}T_{iv}},
\end{equation}
where $\mathrm{diag}(\bA)$ is a diagonal matrix taking the diagonal elements in a square matrix $\bA$ and $\det(\bA)$ is the determinant of $\bA$. When $\hat{\Gamma}^{(k)}$ diagnolizes all $\hat{\Sigma}_{iv}$'s, $\mathrm{DfD}(\hat{\Gamma}^{(k)})=1$, otherwise, it is greater than $1$. In practice, one can set a threshold, such as $\mathrm{DfD}(\hat{\Gamma}^{(k)})\leq 2$, to determine $k$ or stop the procedure before a sudden jump in the metric occurs.
%==========================================

%==========================================
\subsection{Inference}
\label{sub:inference}

To draw inference on the parameters, a bootstrap procedure is employed, which has been proven to yield satisfactory results with small sample size under minimal assumptions~\citep{van2008resampling}. In this study, we propose a nonparametric bootstrapping procedure. Consider the case that the number of visits is small, for example, no more than five visits. We resample the subjects with replacement and keep the visit data for each subject intact~\citep{davison1997bootstrap,van2008resampling,goldstein2011bootstrapping}. Theoretical and simulation studies show that bootstrapping on the highest level offers better performance and a more accurate reflection of the original sample information~\citep{ren2010nonparametric}. Using all the samples, an estimate of $\bgamma$ is first obtained, denoted as $\hat{\bgamma}$. For the $b$th replication, the subjects are resampled with replacement. For each subject, the data from all visits and all the observations at each visit are used to estimate the model parameters, $(\beta_{0i},\bbeta_{1},\beta_{0},\sigma^{2})$, by setting $\bgamma=\hat{\bgamma}$. This procedure is repeated for $B$ times. Confidence intervals of the coefficient parameters are then calculated using either the percentile or bias-corrected approach~\citep{efron1987better}. Here, we focus on the inference of the model coefficient parameters. Inference on $\bgamma$ through a bootstrap requires a matching procedure of the estimate from each bootstrap sample. This is one of our future research direction. 
%==========================================

%==========================================
\subsection{Asymptotic properties}
\label{sub:asmp}

Let $(\bgamma^{*},\bbeta_{1}^{*},\beta_{0}^{*},\sigma^{2*})$ denote the true parameters. We discuss the asymptotic properties of the proposed estimator under two scenarios: (i) $p<\min_{i,v}T_{iv}$ and fixed and (ii) $p>\max_{i,v}T_{iv}$. When $p<\min_{i,v}T_{iv}$ and fixed, one can replace $\hat{\Sigma}_{iv}$ with the sample covariance matrix, $\bS_{iv}$, in~\eqref{eq:hll}. Minimizing~\eqref{eq:hll} is then equivalent to maximizing the hierarchical likelihood function. Thus, with $\bgamma=\bgamma^{*}$, the estimator of $(\bbeta_{1}^{*},\beta_{0}^{*},\sigma^{2*})$ is consistent~\citep{andersen1970asymptotic,lee1996hierarchical}.
\begin{theorem}\label{thm:asmp_sp}
  Assume $p<\min_{i,v}T_{iv}$ and $p$ is fixed. Let $M_{n}=\sum_{i}\sum_{v}T_{iv}$, $T=\min_{i,v}T_{iv}$, and $V=\min_{i}V_{i}$. Set $\hat{\Sigma}_{i}=\bS_{i}$ in~\eqref{eq:hll}, where $\bS_{i}=\sum_{t=1}^{T_{iv}}\by_{iv}\by_{iv}^\top/T_{iv}$ is the sample covariance matrix. For $\bbeta_{1}$,
  \begin{equation}
    \sqrt{M_{n}}\left(\hat{\bbeta}_{1}-\bbeta_{1}^{*}\right)\overset{\mathcal{D}}{\longrightarrow}\mathcal{N}(\boldsymbol{\mathrm{0}},2\bQ^{-1}), \quad \text{as } n,T\rightarrow \infty,
  \end{equation}
  where $\sum_{i}\sum_{v}(\bx_{iv}\bx_{iv}^\top/nV_{i})\rightarrow \bQ$; and for $\beta_{0}$ and $\sigma^{2}$,
  \begin{eqnarray}
    && \hat{\beta}_{0}\overset{\mathcal{P}}{\longrightarrow}\beta_{0}^{*}, \quad \text{as } n,T\rightarrow\infty, \\
    && \hat{\sigma}^{2}\overset{\mathcal{P}}{\longrightarrow}\sigma^{2*}, \quad \text{as } n,V,T\rightarrow\infty.
  \end{eqnarray}
\end{theorem}
When assuming that all the covariance matrices have the same set of eigenvectors, the estimator of $\bgamma$ by Algorithm~\ref{alg:covreg} (denoted as $\hat{\bgamma}$) is a consistent estimator based on the theory of maximum likelihood estimator. Thus, the estimator of $(\bbeta_{1}^{*},\beta_{0}^{*},\sigma^{2*})$ under $\hat{\bgamma}$ is also consistent.

Now, we discuss the case of $p>\max_{i,v}T_{iv}$. Under this scenario, one should replace $\hat{\Sigma}_{iv}$ with the proposed shrinkage estimator rather than the sample covariance matrix as $\bS_{iv}$ is rank deficient. This is a generalization of the proposal in \citet{zhao2021principal} to a longitudinal setting, thus we leave all the detailed discussion, such as the imposed assumptions, to Appendix Section~\ref{appendix:sec:asmp_lp} and only present the main result here. Let $\bar{\bS}=\sum_{i=1}^{n}\sum_{v=1}^{V_{i}}T_{iv}\bS_{iv}/\sum_{i=1}^{n}\sum_{v=1}^{V_{i}}T_{iv}$ denote the average sample covariance matrix over subjects and visits. Under Assumptions A2 and A5, $\bar{\bS}$ is guaranteed to be positive definite and the eigenvectors of $\bar{\bS}$ are consistent estimators~\citep{anderson1963asymptotic}. Taking the eigenvectors of $\bar{\bS}$ as the initial values of $\bgamma$, the estimators from Algorithm~\ref{alg:covreg} are consistent.
\begin{theorem}
  Under Assumptions A1--A5 (in Appendix Section~\ref{appendix:sec:asmp_lp}), as $n,T\rightarrow\infty$, the estimator of $(\bgamma^{*},\bbeta_{1}^{*},\beta_{0}^{*})$ obtained by Algorithm~\ref{alg:covreg} is asymptotically consistent. In addition, as $n,V,T\rightarrow\infty$, the estimator of $\sigma^{2*}$ is asymptotically consistent.
\end{theorem}
%==========================================

%%%%%%%%%%%%%%%%%%%%%%%%%%%%%%%%%%%%%%%%%%%%%%%%%%%%%%%%%%

%%%%%%%%%%%%%%%%%%%%%%%%%%%%%%%%%%%%%%%%%%%%%%%%%%%%%%%%%%
% Simulation
%%%%%%%%%%%%%%%%%%%%%%%%%%%%%%%%%%%%%%%%%%%%%%%%%%%%%%%%%%
\section{Simulation Study}
\label{sec:sim}

In this section, we present the performance of the proposed method through simulation studies. Comparisons with other competing methods under the cross-sectional setting were studied for the case of $p<T_{\min}$ and $p>T_{\max}$ in \citet{zhao2021covariate} and \citet{zhao2021principal}, respectively. In this manuscript, we compare the proposed longitudinal covariate assisted principal regression approach (denoted as LCAP) with a longitudinal approach derived from the approach in \citet{zhao2021principal} (denoted as CAP-mix). The CAP-mix approach includes three steps: (1) apply the covariance regression model in \citet{zhao2021principal} to the data collected at the first visit to estimate the projection and shrinkage parameters; (2) use the estimates to acquire the log-transformed scores, $\log(\hat{\bgamma}^\top\hat{\Sigma}_{iv}\hat{\bgamma})$, for each subject at each visit; and (3) fit these scores in a mixed effects model to yield the estimate and inference of $\bbeta$.

In this longitudinal simulation study, for subject $i$ at visit $v$, the covariance matrices are generated using the eigendecomposition $\Sigma_{iv}=\Pi\Lambda_{iv}\Pi^\top$, where $\Pi=(\bpi_{1},\dots,\bpi_{p})$ is an orthonormal matrix in $\mathbb{R}^{p\times p}$ and $\Lambda_{iv}=\mathrm{diag}\{\lambda_{iv1},\dots,\lambda_{ivp}\}$ is a diagonal matrix. A case of two covariates is considered (thus $q=3$), a binary outcome generated from a Bernoulli distribution with probability $0.5$ to be one and a continuous outcome generated from a normal distribution with mean zero and variance $0.5^2$. The intercept coefficient, $\beta_{0}$, exponentially decays from $3$ to $-3$, and $\beta_{0i}$ is generated from a normal distribution with mean $\beta_{0}$ and variance $\sigma^{2}=0.1^{2}$. Two dimensions, D2 and D4, are chosen to be related to the covariates. For D2, $\beta_{1}=-0.5$ and $\beta_{2}=0.5$; for D4, $\beta_{1}=0.5$ and $\beta_{2}=-0.25$; and for the rest, $\beta_{1}=\beta_{2}=0$.
For both LCAP and CAP-mix methods, the number of components is chosen with $\mathrm{DfD}(\hat{\bGamma}^{(k)})< 2$. The performance of estimating the number of components under this threshold has been studied in \citet{zhao2021covariate}. Thus, we will not repeat here. In this manuscript, we only present the result of identifying D4 for demonstration. For $\bbeta$, we choose to present the result of estimating $\beta_{2}$ as it has a relatively smaller effect size.

\begin{figure}
  \begin{center}
    \subfloat[Bias of $\hat{\beta}$]{\includegraphics[width=0.3\textwidth]{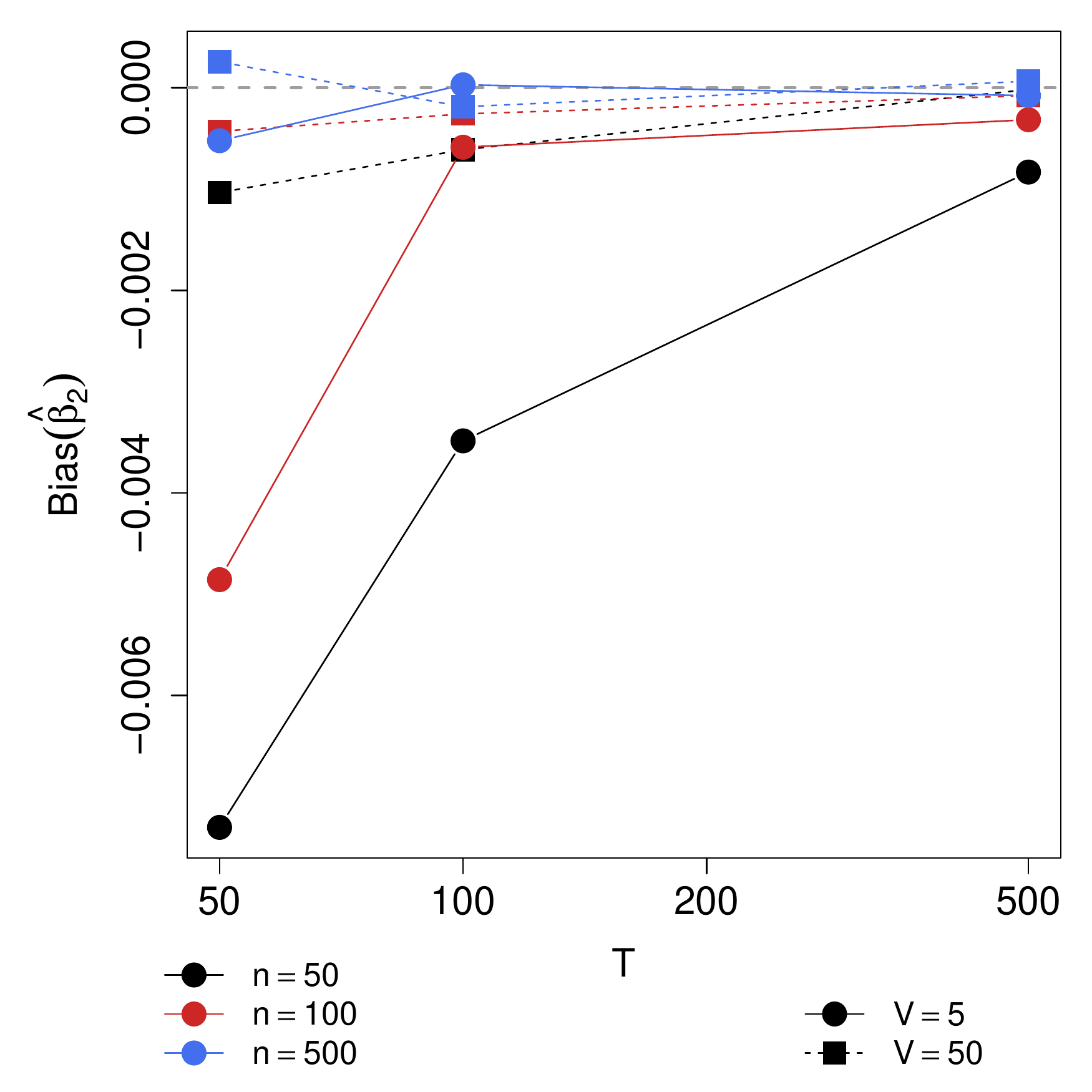}}
    \enskip{}
    \subfloat[MSE of $\hat{\beta}$]{\includegraphics[width=0.3\textwidth]{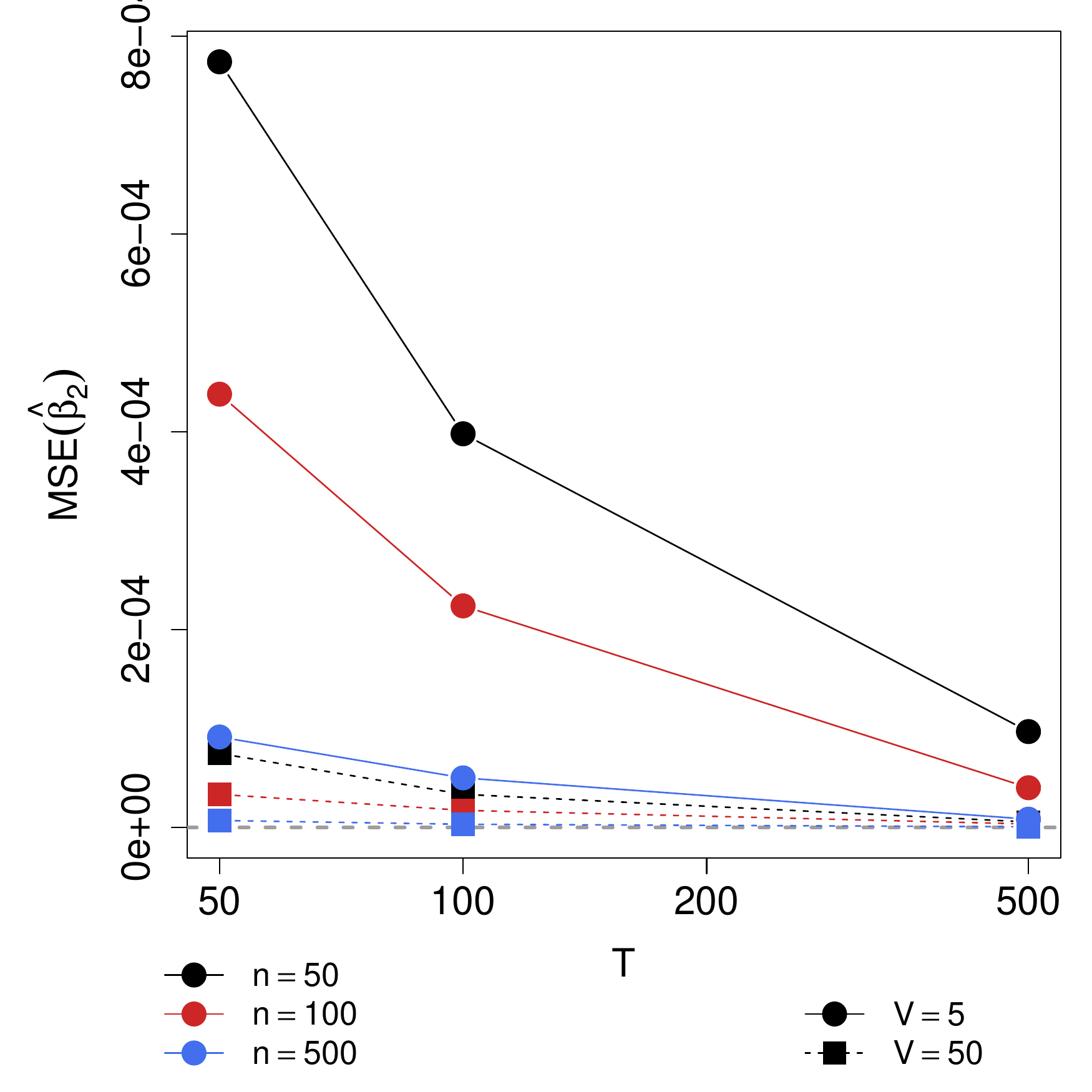}}
    \enskip{}
    \subfloat[CP of $\hat{\beta}$]{\includegraphics[width=0.3\textwidth]{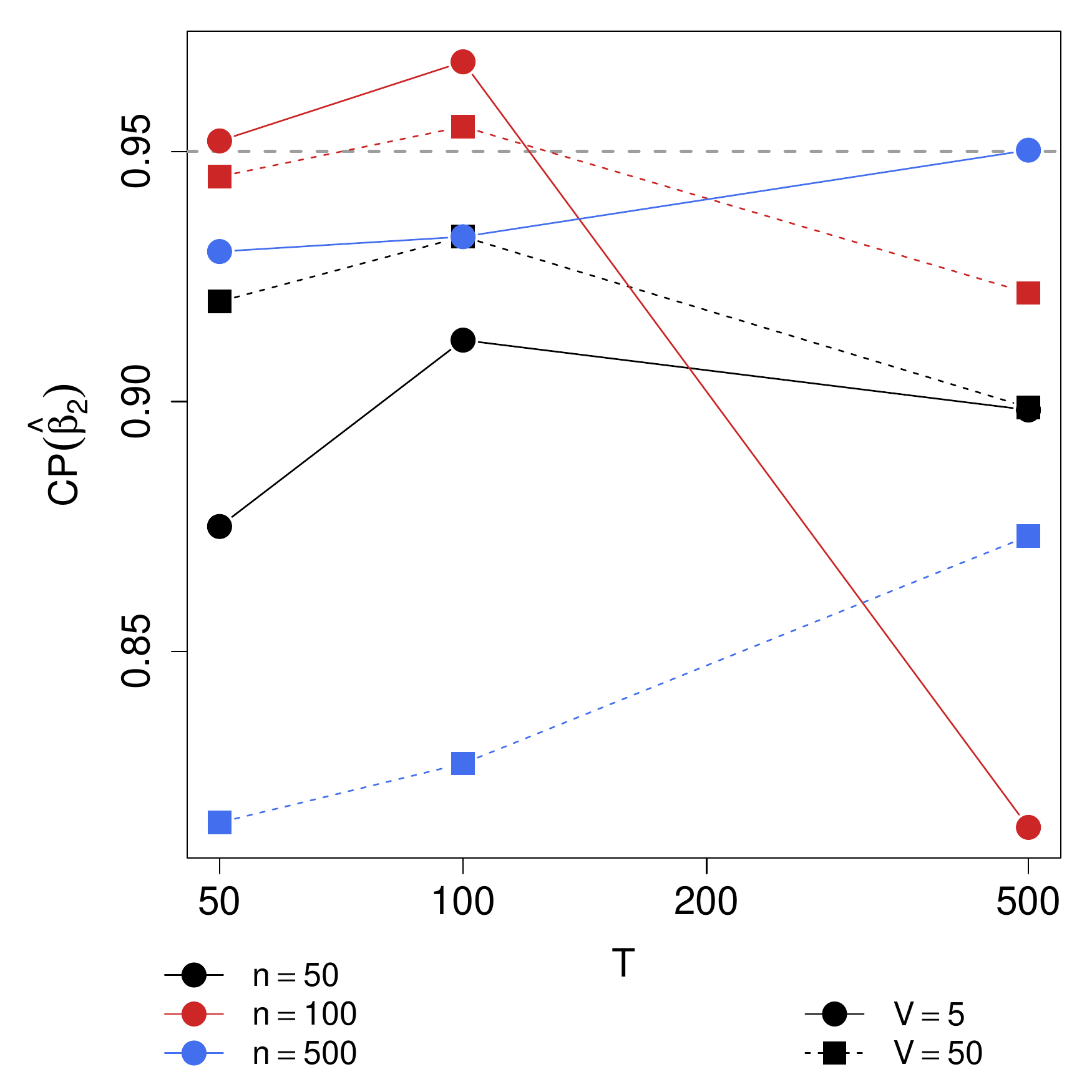}}

    \subfloat[Similarity metric of $\hat{\bgamma}$]{\includegraphics[width=0.3\textwidth]{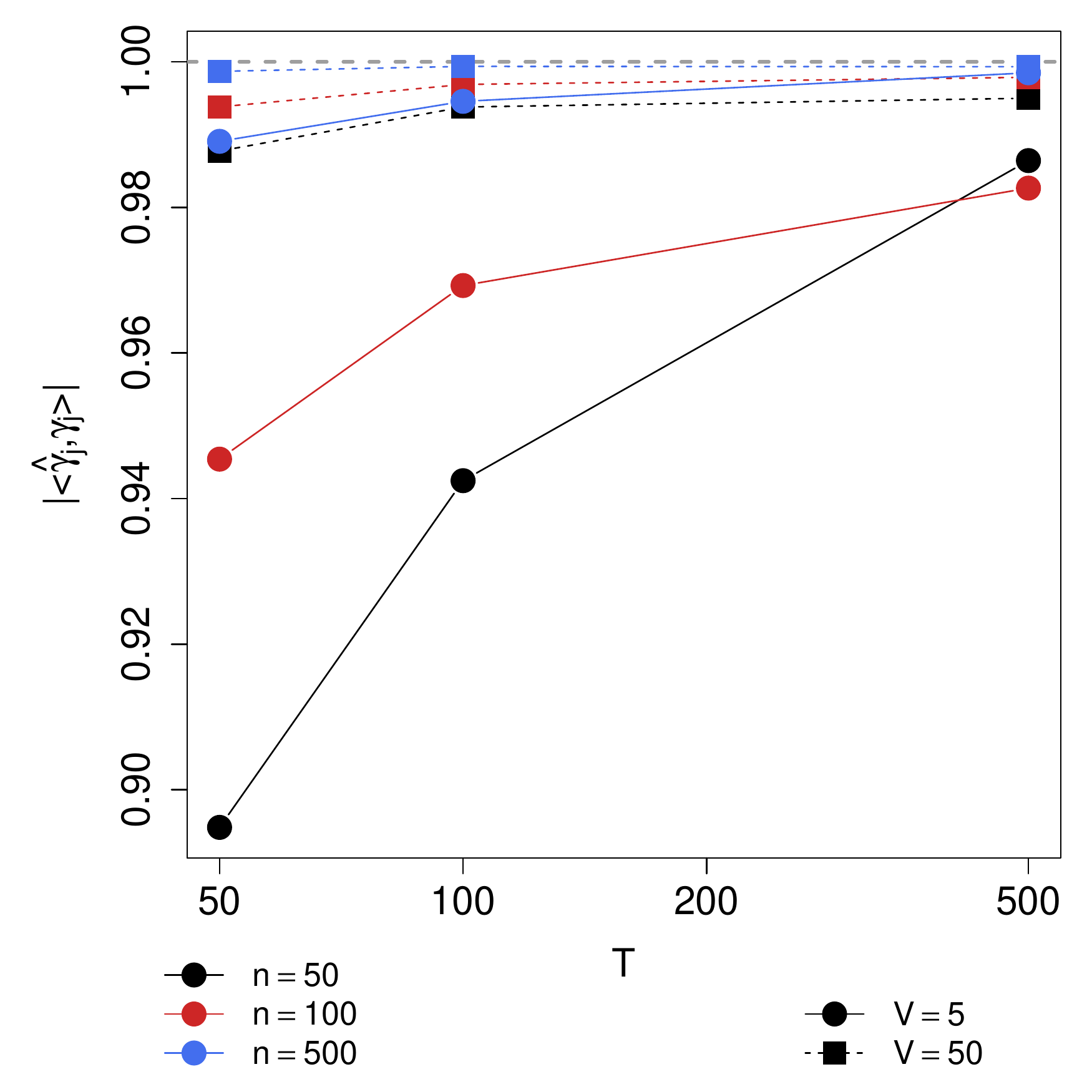}}
    \enskip{}
    \subfloat[Bias of $\hat{\sigma}^{2}$]{\includegraphics[width=0.3\textwidth]{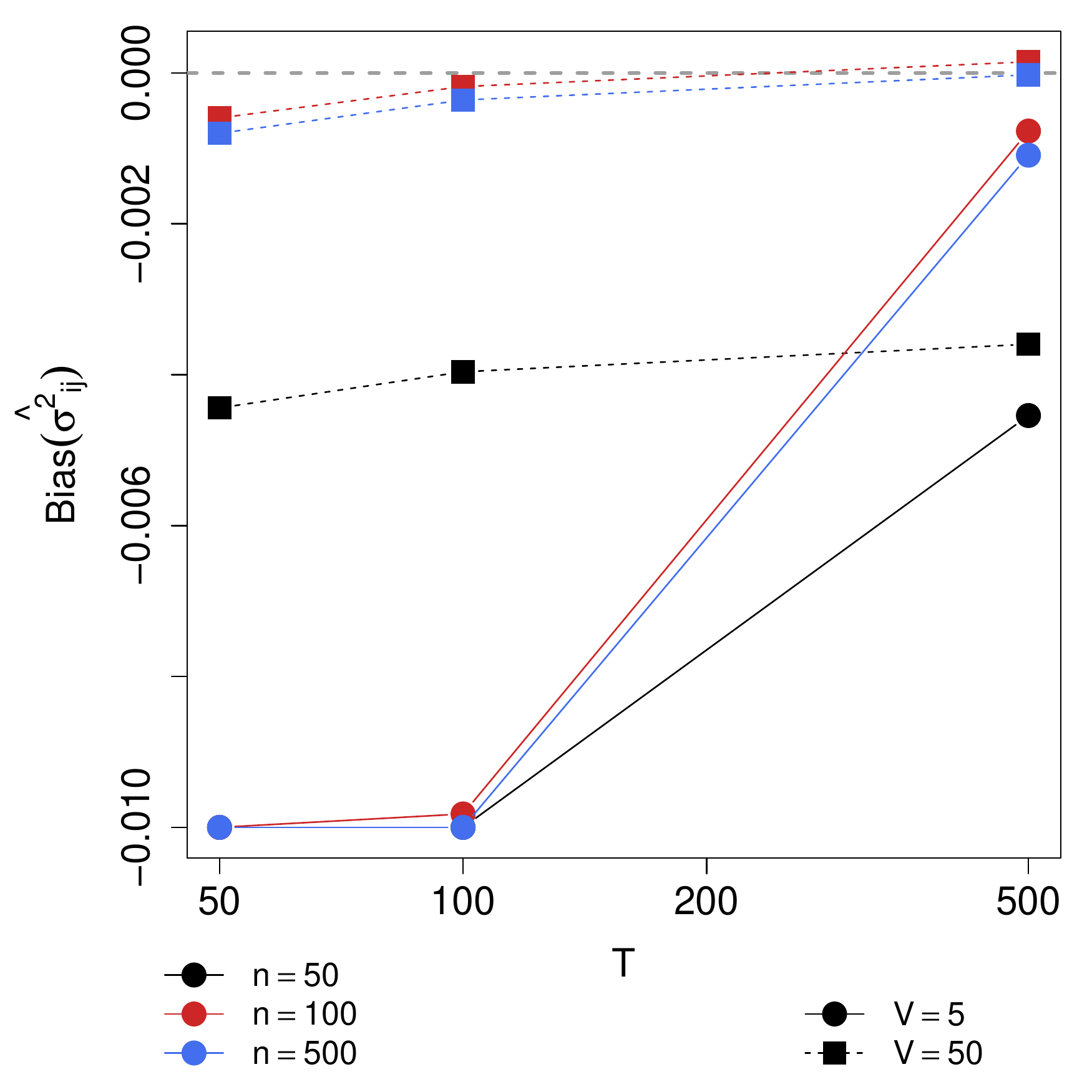}}
    \enskip{}
    \subfloat[MSE of $\hat{\sigma}^{2}$]{\includegraphics[width=0.3\textwidth]{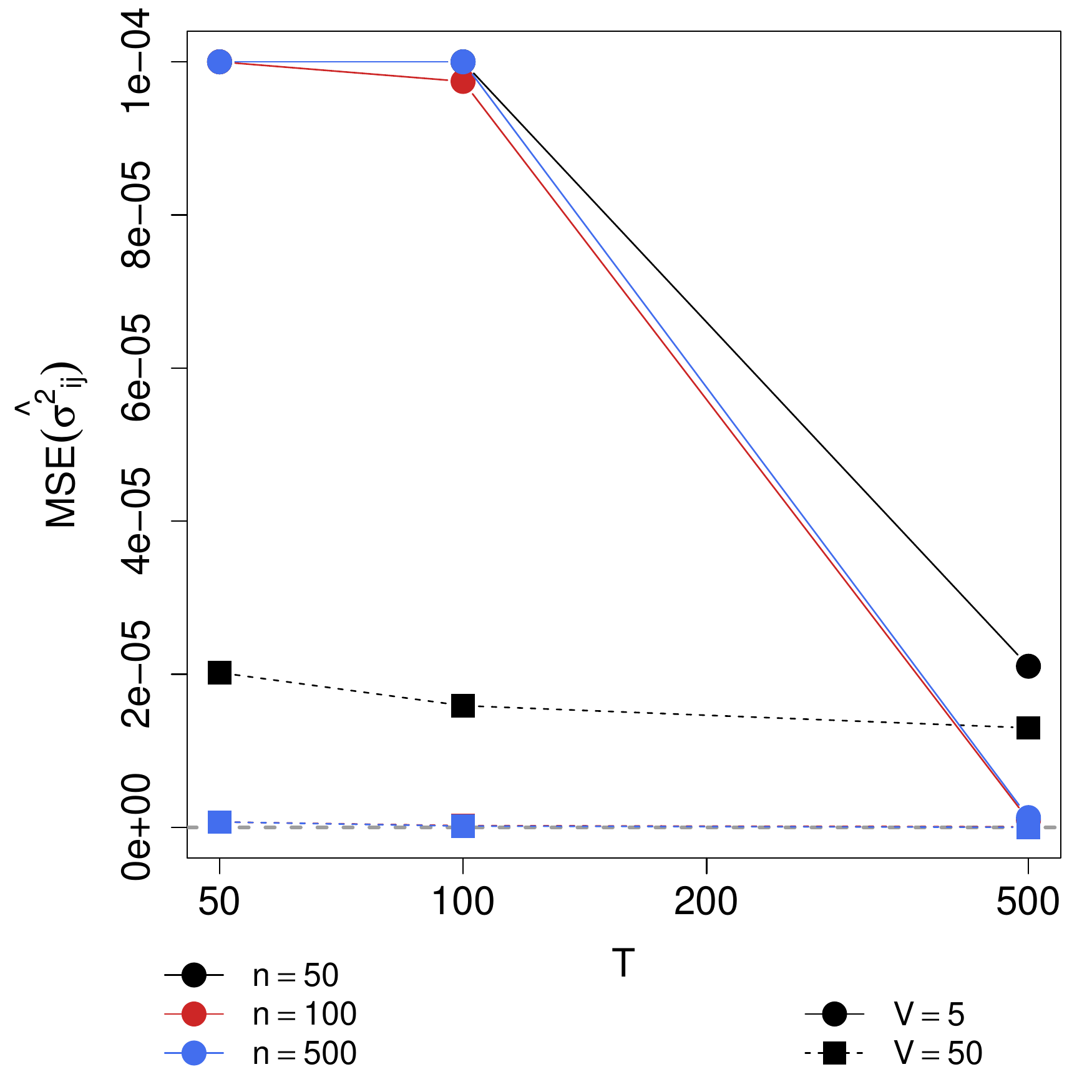}}
  \end{center}
  \caption{\label{fig:sim_p20}Estimation performance in estimating the parameters in the simulation study. For $\hat{\beta}$, (a) bias, (b) mean squared error (MSE) and (c) coverage probability (CP) from $500$ bootstrap samples. For $\hat{\bgamma}$, (e) similarity to $\bpi_{j}$. For $\sigma^{2}$, (e) bias and (f) MSE. Data dimension $p=20$. Sample sizes vary from $n=50,100,500$, $V_{i}=V=5,50$, and $T_{iv}=T=50,100,500$.}
\end{figure}

We first consider a case of $p=20$ and set $n=50,100,500$, $V_{i}=V=5,50$, and $T_{iv}=T=50,100,500$ to examine the finite sample performance. The results are presented in Figure~\ref{fig:sim_p20}. From the figures, as $n$ and $T$ increase, the estimate of $\beta$ and $\bgamma$ converge to the truth. For the variance of the random intercept, $\sigma^{2}$, as $V$ increases, the performance improves and the estimate converges to the truth consisting with the theoretical results. 
We compare the performance of the proposed LCAP approach with the CAP-mix approach in the setting with a higher dimension, $p=100$. Table~\ref{table:sim_p100} presents the results. Though LCAP achieves slightly higher bias under lower sample size with $V=5$ and $T_{i}=T=50$, when either of $V$ and $T$ increases, the performance of LCAP improves that it yields lower bias and mean squared error (MSE) in estimating $\beta$ and higher correlation to the truth in estimating $\bgamma$. In addition, as $V$ increases, the performance of CAP-mix does not improve and the coverage probability of $\beta$ estimate decreases. Thus, for longitudinal data, the proposed LCAP approach is a more appropriate choice. In \citet{zhao2021principal}, the robustness of the methods was examined under two types of model misspecification, in $\bbeta$ and in $\bgamma$. The conclusions are expected be generalizable to the longitudinal setting. Thus, we will not repeat the evaluation of the robustness in this study.

\begin{table}
  \caption{\label{table:sim_p100}Bias, mean squared error (MSE), and coverage probability (CP) in estimate $\beta$, the similarity of $\hat{\bgamma}$ to $\bpi_{j}$, and the MSE in estimating the eigenvalues in the simulation study. Data dimension $p=100$, sample size $n=50$ and $T_{iv}=T=50,500$, and the number of visits $V=5,50$. For CAP-mix, the CP is calculated from the mixed effects model; for LCAP, the CP is calculated from $500$ bootstrap samples.}
  \begin{center}
    \begin{tabular}{l l l r c c c c c c}
      \hline
      & & & \multicolumn{3}{c}{$\hat{\beta}$} && \multicolumn{1}{c}{$\hat{\bgamma}$} && \multicolumn{1}{c}{$\hat{\lambda}_{ij}$} \\
      \cline{4-6}\cline{8-8}\cline{10-10}
      \multicolumn{1}{c}{\multirow{-2}{*}{$V$}} & \multicolumn{1}{c}{\multirow{-2}{*}{$T$}} & \multicolumn{1}{c}{\multirow{-2}{*}{Method}} & \multicolumn{1}{c}{Bias} & \multicolumn{1}{c}{MSE ($\times 10^{-3}$)} & \multicolumn{1}{c}{CP} && \multicolumn{1}{c}{$|\langle \hat{\bgamma},\bpi_{j} \rangle|$ (SE)} && \multicolumn{1}{c}{MSE} \\
      \hline
      & & CAP-mix & $0.006$ & $0.749$ & $0.950$ && $0.842$ ($0.043$) && $0.188$ \\
      & \multirow{-2}{*}{$50$} & LCAP & $-0.011$ & $0.864$ & $0.761$ && $0.618$ ($0.053$) && $0.192$ \\
      \cline{2-10}
      & & CAP-mix & $0.007$ & $0.083$ & $1.000$ && $0.834$ ($0.027$) && $0.162$ \\
      \multirow{-4}{*}{$5$} & \multirow{-2}{*}{$500$} & LCAP & $-0.002$ & $0.071$ & $0.766$ && $0.938$ ($0.011$) && $0.141$ \\
      \hline
      & & CAP-mix & $0.020$ & $0.524$ & $0.376$ && $0.833$ ($0.049$) && $0.207$ \\
      & \multirow{-2}{*}{$50$} & LCAP & $-0.004$ & $0.096$ & $0.766$ && $0.938$ ($0.012$) && $0.163$ \\
      \cline{2-10}
      & & CAP-mix & $0.008$ & $0.078$ & $0.235$ && $0.846$ ($0.025$) && $0.143$ \\
      \multirow{-4}{*}{$50$} & \multirow{-2}{*}{$500$} & LCAP & $0.001$ & $0.006$ & $0.946$ && $0.993$ ($0.001$) && $0.129$ \\
      \hline
    \end{tabular}
  \end{center}
\end{table}

%%%%%%%%%%%%%%%%%%%%%%%%%%%%%%%%%%%%%%%%%%%%%%%%%%%%%%%%%%

%%%%%%%%%%%%%%%%%%%%%%%%%%%%%%%%%%%%%%%%%%%%%%%%%%%%%%%%%%
% Real data
%%%%%%%%%%%%%%%%%%%%%%%%%%%%%%%%%%%%%%%%%%%%%%%%%%%%%%%%%%
\section{The Alzheimer's Disease Neuroimaging Initiative Study}
\label{sec:adni}

We apply the proposed approach to the MRI data collected by the Alzheimer's Disease Neuroimaging Initiative (ADNI, \url{adni.loni.usc.edu}).
%----------------------
% ADNI user agreement
% Data used in this study are obtained from the Alzheimer's Disease Neuroimaging Initiative (ADNI) database (\url{adni.loni.usc.edu}). 
The ADNI study was launched in 2003 as a public-private partnership, led by Principal Investigator Michael W. Weiner, MD. The primary goal of ADNI has been to test whether serial magnetic resonance imaging (MRI), positron emission tomography (PET), other biological markers, and clinical and neuropsychological assessments can be combined to measure the progression of mild cognitive impairment (MCI) and early Alzheimer's disease (AD). 
%----------------------

In the study, resting-state fMRI data were collected at multiple visits. In this study, we focus on the first $V=5$ visits (initial screening, 3-month, 6-month, 1-year, and 2-year visit) for sample size consideration, where $n=78$ subjects with at least three continuous visits are studied (11 subjects with 3 visits, 29 with 4 visits, and 38 with 5 visits). At the initial screening, $14$ are cognitive normal (CN) subjects ($6$ Female), $41$ diagnosed with MCI ($16$ Female), and $23$ diagnosed with AD ($11$ Female).
The fMRI time courses are extracted from $p=75$ brain regions (60 cortical and 15 subcortical regions), which are grouped into ten functional modules, using the Harvard-Oxford Atlas in FSL~\citep{smith2004advances}. For each time course, a subsample is taken with an effective sample size of $T_{iv}=T=67$ to remove the temporal dependence. Denoting the subsampled data as $\by_{ivt}$ (for $t=1,\dots,T$, $v=1,\dots,V_{i}$, and $i=1,\dots,n$), it is assumed that the data follow a multivariate normal distribution with mean zero and covariance matrix $\Sigma_{iv}$. $\Sigma_{iv}$ reveals the architecture of brain functional connectivity of subject $i$ at visit $v$.
In AD research, the effects of sex on dementia is currently of intense investigation. Cumulative evidence suggests sex-specific patterns of disease manifestation and the existence of sex-related difference in the rates of cognitive decline and brain atrophy. Thus, sex is a crucial factor of disease heterogeneity. After diagnosed with MCI or AD, the rate of cognitive decline and brain atrophy was found faster in female than in male~\citep{hua2010sex,skup2011sex,holland2013higher,lin2015marked,tifratene2015progression,ardekani2016analysis,gamberger2017identification}.
In this study, we aim to investigate the sex-related difference in brain functional connectivity with the availability of longitudinal fMRI data. Thus, in the proposed longitudinal regression model, disease diagnosis, sex, and their interaction, as well as age, are entered as the covariates ($\bx_{iv}$'s). 

\begin{figure}
  \begin{center}
    \subfloat[C1]{\includegraphics[width=0.2\textwidth]{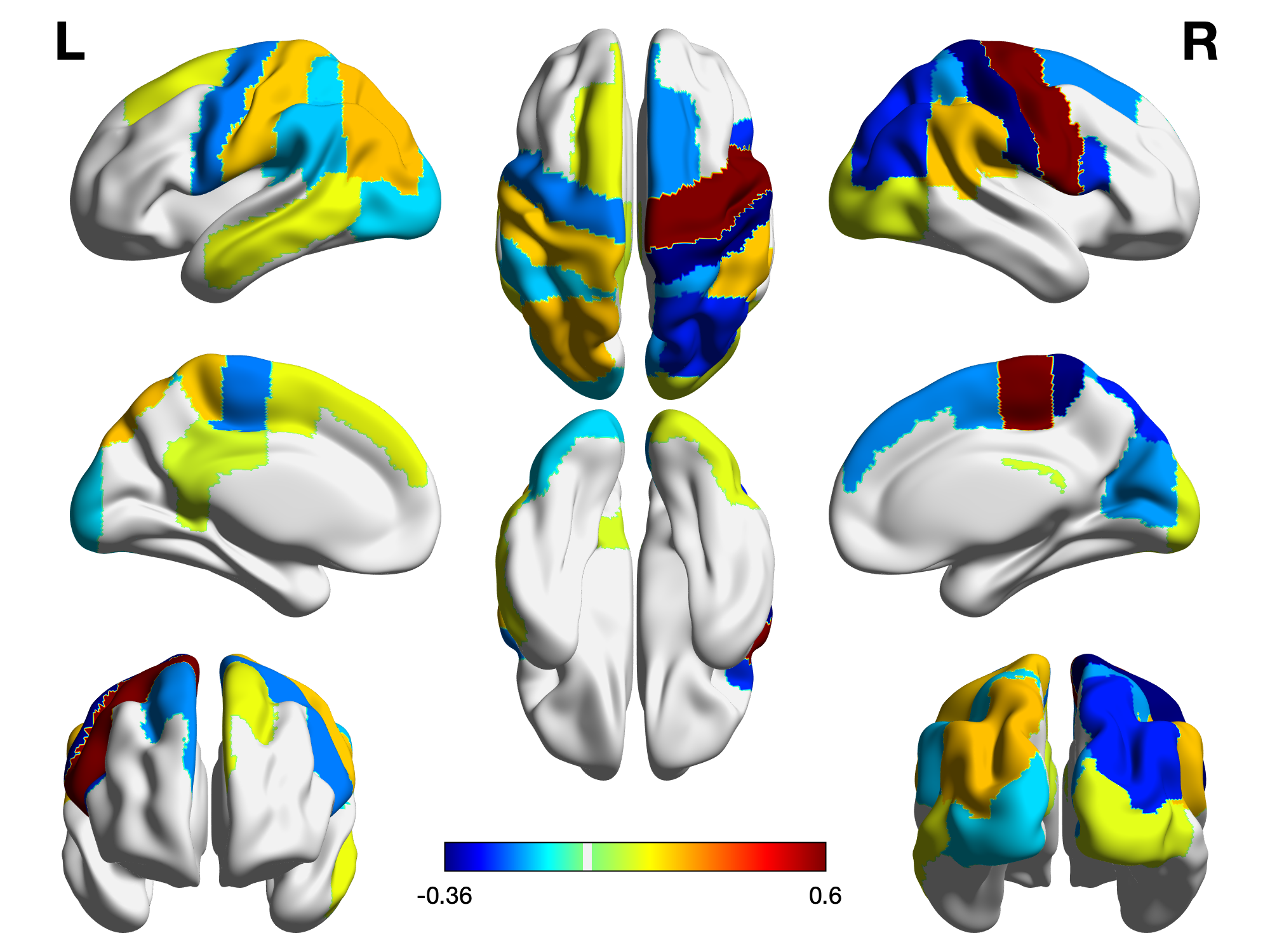}}
    \subfloat[C2]{\includegraphics[width=0.2\textwidth]{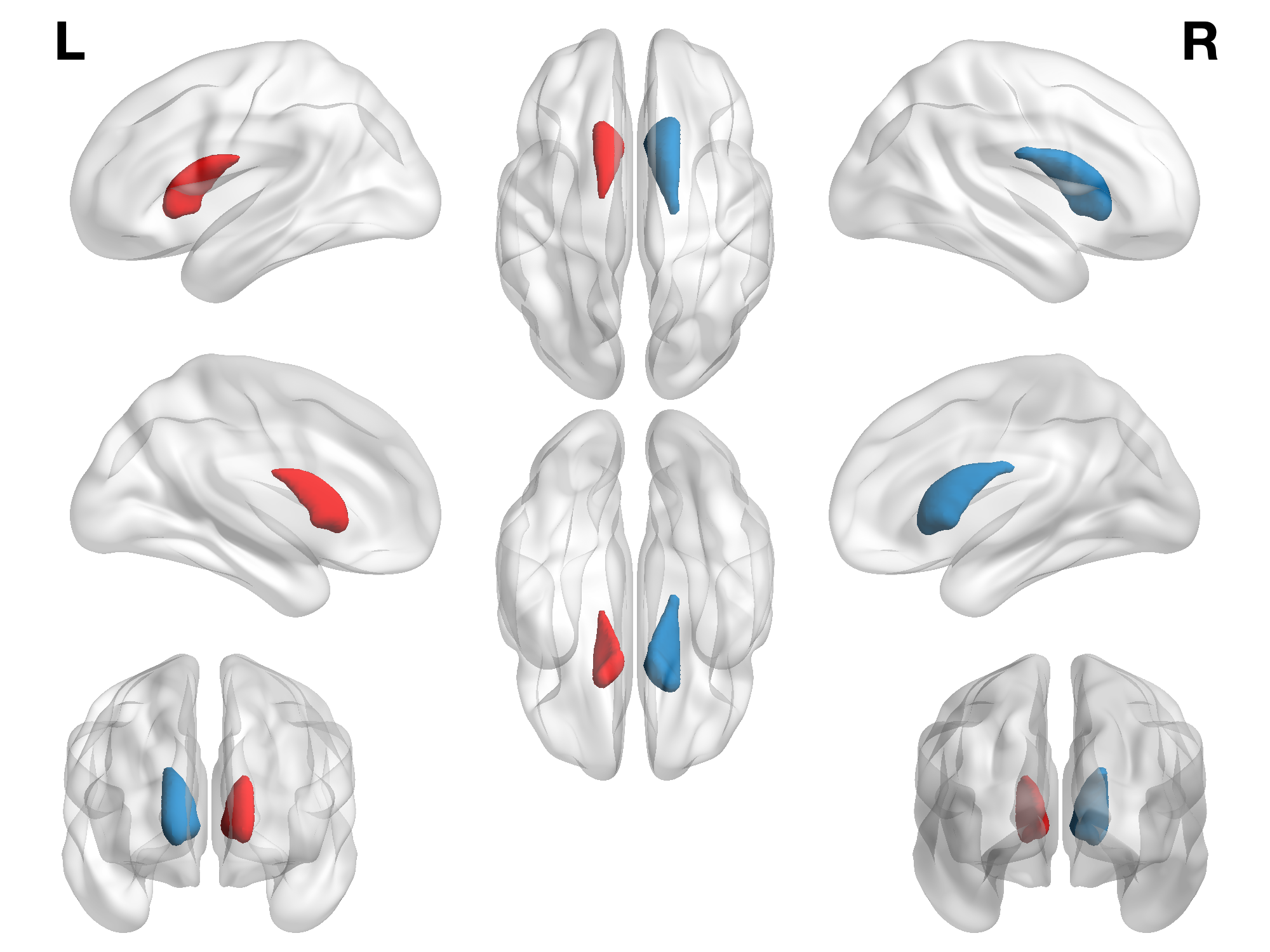}}
    \subfloat[C3]{\includegraphics[width=0.2\textwidth]{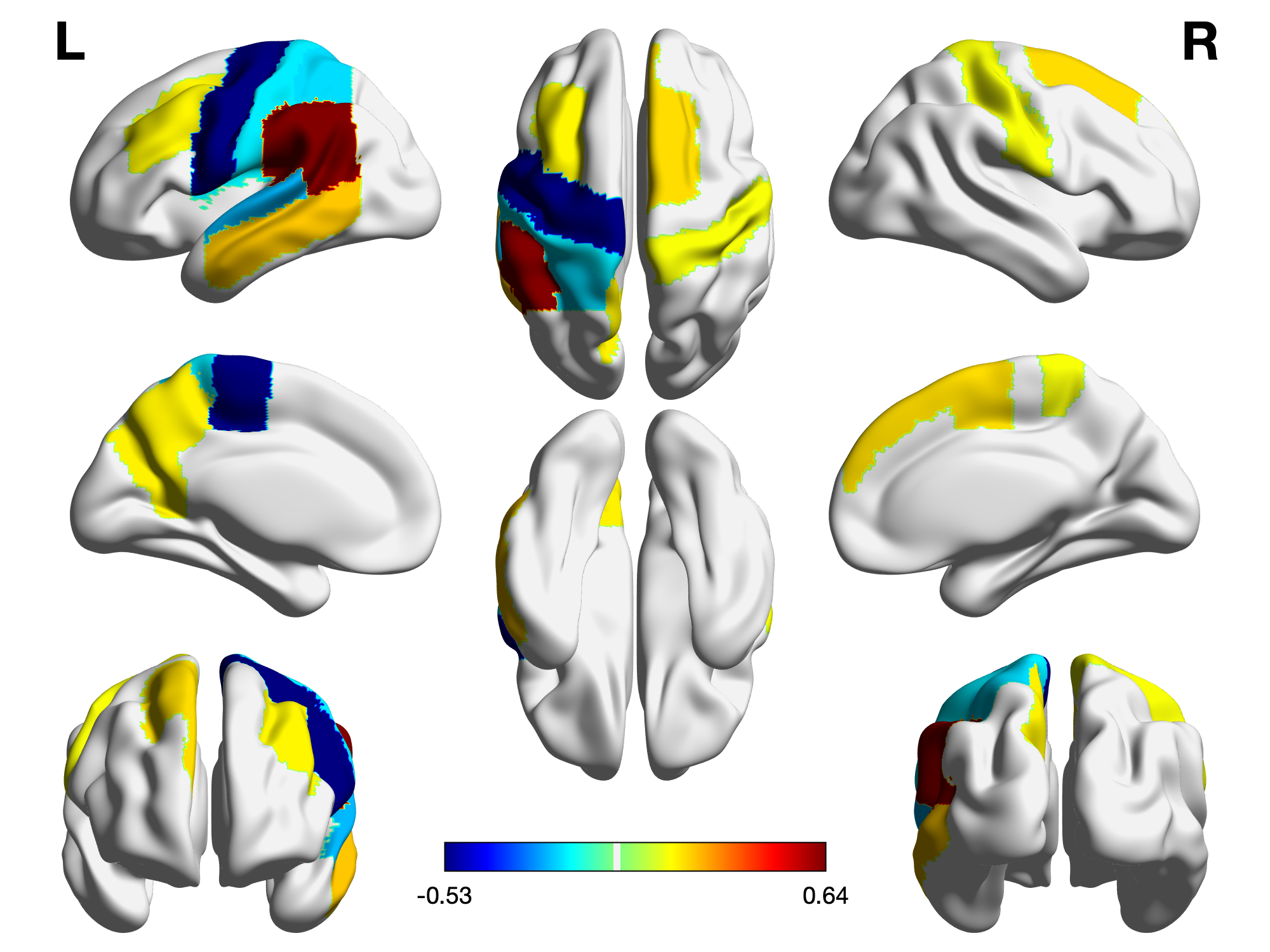}}
    \subfloat[C4]{\includegraphics[width=0.2\textwidth]{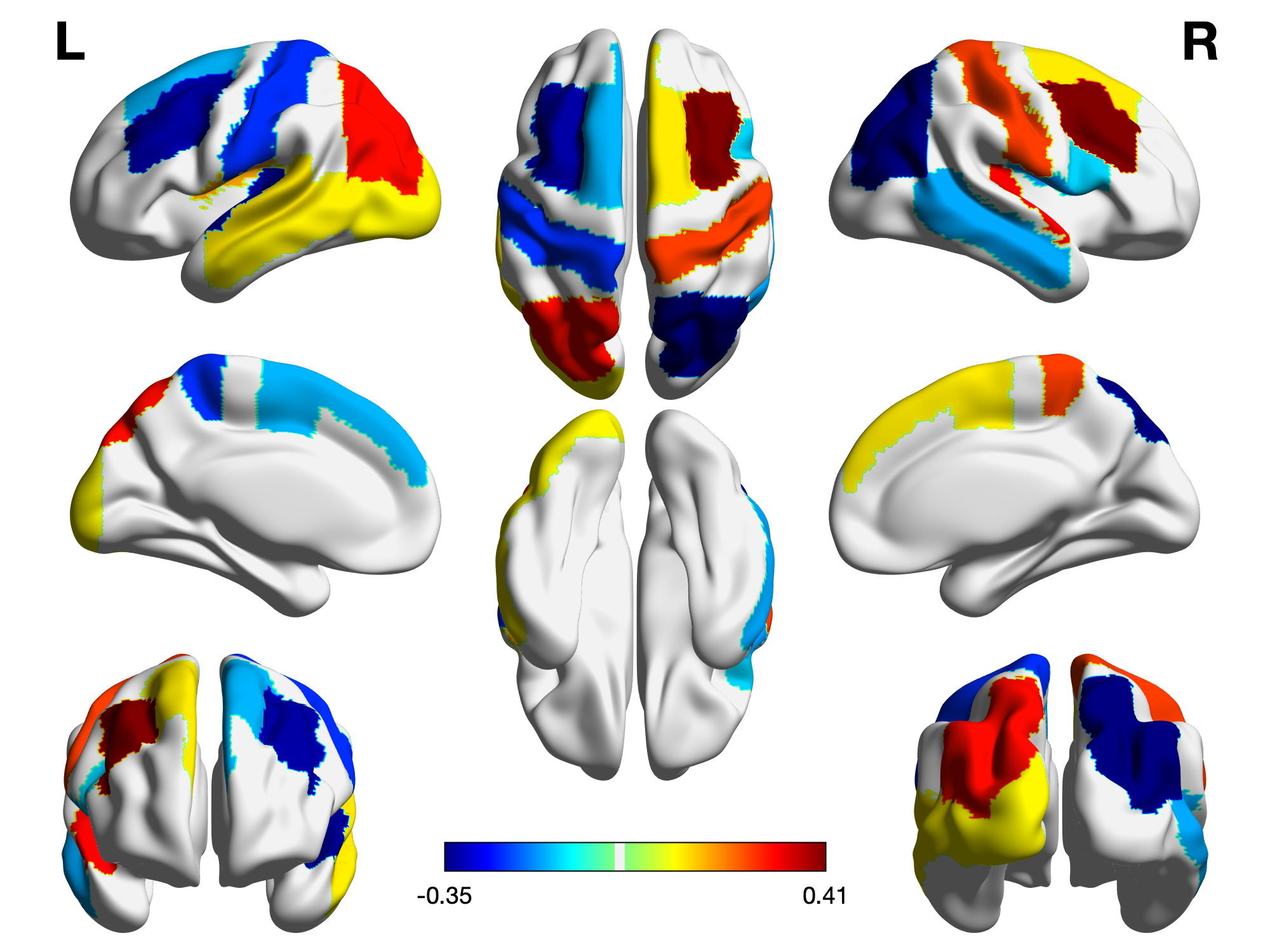}}
    \subfloat[C5]{\includegraphics[width=0.2\textwidth]{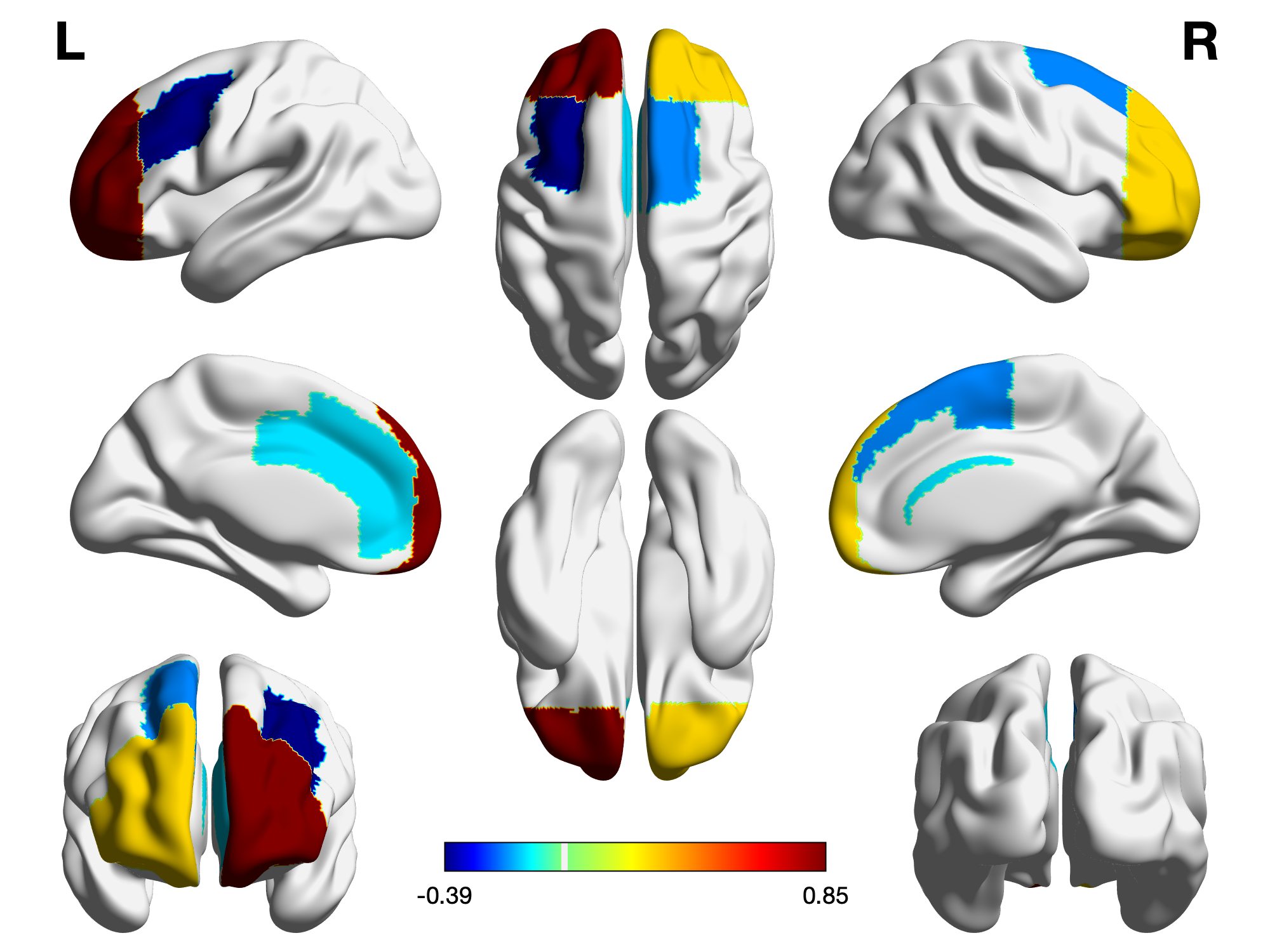}}
  \end{center}
  \caption{\label{fig:adni_brain}The brain map of the five identified components in the ADNI analysis.}
\end{figure}

The proposed approach identifies five orthogonal components, denoted as C1--C5, using the average deviation from diagonality metric setting the threshold at two. With an interaction of diagnosis and sex, pair-wise subgroup comparisons are conducted and the results are presented in Table~\ref{table:adni}. Table~\ref{appendix:table:adni_sigma2} in Appendix Section~\ref{appendix:sec:adni} presents the estimated within-subject variation ($\sigma^{2}$) of each component.
For C1 and C5, the functional connectivity is significantly different between MCI/AD and CN in both females and males. C2 suggests a difference for all pairwise group comparisons in females and C3 suggests a difference comparing MCI and AD to CN. For males, C2 and C4 demonstrate a difference between MCI and CN; and C5 demonstrates a difference in all pairwise group comparisons. In components C1, C2, C4, and C5, a significant difference between males and females is observed in the CN group; and the sex difference in the AD group is observed in C4 and C5. In C1, C2, and C3, it is also found that functional connectivity is significantly associated with age. As age increases, the integrity of the network connectivity decreases. 
Figure~\ref{appendix:fig:adni_scoreMean} in the supplementary materials presents the longitudinal trajectory of each component's connectivity (represented by $\log(\hat{\bgamma}^\top\hat{\Sigma}_{iv}\hat{\bgamma})$) for each diagnosis-sex subgroup over the five visits. For all the components, as time progresses, the level of connectivity decreases. Subgroup differences are observed and are consistent with the results in Table~\ref{table:adni}.
Figure~\ref{appendix:fig:adni_loading} presents the sparsified loading profile of these five components, where the modular information of the brain regions is incorporated. Figure~\ref{fig:adni_brain} shows the regions in brain maps and Table~\ref{table:adni_network} shows the brain networks covered by each component. Applying the identified components to data collected at each individual visit, Table~\ref{appendix:table:adni_cs} in the supplementary materials shows the significance of the comparisons for visits 1--3. From the table, for those significant results, the direction of the association is consistent across visits. When comparing the two tables (Tables~\ref{table:adni} and~\ref{appendix:table:adni_cs}), it suggests that the longitudinal approach improves the statistical power of identifying the associations after integrating the data from multiple visits.

% AD/MCI vs CN
Consistent decrease in DMN connectivity among MCI and AD cohorts has been reported and the findings are robust to the choice of analytical approaches~\citep{badhwar2017resting}.
Regions, such as precentral and postcentral gyri, have been verified to be related to working memory demonstrating the difference in functional connectivity between MCD/AD patients and normal controls~\citep{tomasi2011methylphenidate}. 
% Sex
The study of sex differences in brain functional connectivity using fMRI is limited in AD research. In a recent cross-sectional preAD study, functional connectivity in the DMN was found to be lower in males than in females~\citep{cavedo2018sex}. In a study comparing amnestic MCI and AD with healthy controls, a difference in right caudate nucleus atrophy was observed in females only. Alterations of functional connectivity in the somato-motor, dorsal, and ventral attention networks were identified in male patients~\citep{li2021sex}.
In summary, our findings are in line with existing knowledge about AD.

\begin{table}
  \caption{\label{table:adni}Comparison between diagnosis groups for females and males and comparison between males and females for each diagnosis group (CN, MCI, and AD) of the five identified components in the ADNI analysis. The numbers in the parentheses are the $95\%$ confidence intervals obtained from $500$ bootstrap samples.}
  \begin{center}
    \resizebox{\textwidth}{!}{
    \begin{tabular}{l l r r r r r}
      \hline
      \multicolumn{1}{c}{Comparison} & \multicolumn{1}{c}{Group} & \multicolumn{1}{c}{C1} & \multicolumn{1}{c}{C2} & \multicolumn{1}{c}{C3} & \multicolumn{1}{c}{C4} & \multicolumn{1}{c}{C5} \\
      \hline
      & Female & $\mathbf{0.48~(~~0.35, ~~0.61)}$ & $\mathbf{0.72~(~~0.51, ~~0.94)}$ & $\mathbf{0.39~(~~0.25, ~~0.53)}$ & $0.03~(-0.05, ~~~0.12)$ & $\mathbf{0.37~(~~0.27, ~~0.46)}$ \\
      \multirow{-2}{*}{MCI$-$CN} & Male & $\mathbf{0.21~(~~0.10, ~~0.32)}$ & $\mathbf{-0.57~(-0.91, -0.23)}$ & $0.21~(-0.04, ~~~0.45)$ & $\mathbf{0.46~(~~0.08, ~~0.83)}$ & $\mathbf{0.55~(~~0.35, ~~0.75)}$ \\
      \hline
      & Female & $\mathbf{0.29~(~~0.08, ~~0.51)}$ & $\mathbf{1.16~(~~0.89, ~~1.43)}$ & $\mathbf{0.30~(~~0.13, ~~0.48)}$ & $0.43~(-0.01, ~~~0.87)$ & $\mathbf{0.40~(~~0.09, ~~0.71)}$ \\
      \multirow{-2}{*}{AD$-$CN} & Male & $\mathbf{0.18~(~~0.02, ~~0.33)}$ & $-0.45~(-0.96, ~~~0.05)$ & $0.14~(-0.18, ~~~0.45)$ & $0.35~(-0.10, ~~~0.81)$ & $\mathbf{0.31~(~~0.04, ~~0.57)}$ \\
      \hline
      & Female & $-0.19~(-0.42, ~~~0.05)$ & $\mathbf{0.12~(~~0.19, ~~0.68)}$ & $-0.09~(-0.22, ~~~0.05)$ & $0.39~(-0.04, ~~~0.83)$ & $0.03~(-0.28, ~~~0.35)$ \\
      \multirow{-2}{*}{AD$-$MCI} & Male & $-0.03~(-0.15, ~~~0.08)$ & $0.11~(-0.24, ~~~0.47)$ & $-0.07~(-0.30, ~~~0.15)$ & $-0.10~(-0.36, ~~~0.16)$ & $\mathbf{-0.24~(-0.44, -0.03)}$ \\
      \hline
      & CN & $\mathbf{0.23~(~~0.01, ~~0.46)}$ & $\mathbf{1.30~(~~0.91, ~~1.69)}$ & $0.20~(-0.09, ~~~0.49)$ & $\mathbf{-0.54~(-1.06, -0.02)}$ & $\mathbf{-0.37~(-0.65, -0.10)}$ \\
      & MCI & $-0.03~(-0.27, ~~~0.21)$ & $0.01~(-0.29, ~~~0.31)$ & $0.02~(-0.23, ~~~0.26)$ & $-0.11~(-0.46, ~~~0.23)$ & $-0.19~(-0.47, ~~~0.09)$ \\
      \multirow{-3}{*}{Male$-$Female} & AD & $0.12~(-0.16, ~~~0.39)$ & $-0.31~(-0.77, ~~~0.14)$ & $0.03~(-0.23, ~~~0.29)$ & $\mathbf{-0.61~(-0.14, -0.07)}$ & $\mathbf{-0.46~(-0.84, -0.08)}$ \\
      \hline
      \multicolumn{2}{l}{Age} & $\mathbf{-0.12~(-0.23, -0.00)}$ & $\mathbf{-0.32~(-0.56, -0.08)}$ & $\mathbf{-0.16~(-0.28, -0.04)}$ & $0.20~(-0.03, ~~~0.43)$ & $0.17~(-0.02, ~~~0.35)$ \\
      \hline
    \end{tabular}
    }
  \end{center}
\end{table}

\begin{table}
  \caption{\label{table:adni_network}Brain networks covered by the five identified components in the ADNI analysis.}
  \begin{center}
    \resizebox{\textwidth}{!}{
    \begin{tabular}{L{15pt} C{40pt} C{40pt} C{40pt} C{40pt} C{40pt} C{40pt} C{40pt} C{60pt} C{60pt}}
      \hline
      & Visual & Somato-motor & Dorsal-attention & Ventral-attention & Limbic-system & Fronto-parietal & DMN & Subcortical & Cerebellum \\
      \hline
      C1 & $\times$ & $\times$ & $\times$ & $\times$ & & & $\times$ & & \\
      C2 & & & & & & & & $\times$ \\
      C3 & & $\times$ & $\times$ & $\times$ & & & $\times$ \\
      C4 & $\times$ & $\times$ & $\times$ & $\times$ & $\times$ & & $\times$ \\
      C5 & & & & & & $\times$ & $\times$ \\
      \hline
    \end{tabular}
    }
  \end{center}
\end{table}

% \begin{itemize}
%   \item \citet{skup2011sex}:
%     \begin{itemize}
%       \item aMCI subjects: difference between males and females in left/right caudate nucleus
%       \item control subjects: difference between males and females in right caudate nucleus
%       \item females (not males): with probable AD and aMCI different from control in right caudate nucleus atrophy
%     \end{itemize}
% \end{itemize}
%%%%%%%%%%%%%%%%%%%%%%%%%%%%%%%%%%%%%%%%%%%%%%%%%%%%%%%%%%

%%%%%%%%%%%%%%%%%%%%%%%%%%%%%%%%%%%%%%%%%%%%%%%%%%%%%%%%%%
% Discussion
%%%%%%%%%%%%%%%%%%%%%%%%%%%%%%%%%%%%%%%%%%%%%%%%%%%%%%%%%%
\section{Discussion}
\label{sec:discussion}

In this manuscript, we propose a longitudinal regression model for covariance matrix outcomes. The proposal considers a multilevel model based on a generalized linear model for covariance matrices to simultaneously identify covariate associated components, estimate model coefficients, and capture the within-subject variation in the covariance matrices. Under the normality assumption, a hierarchical likelihood-based approach is introduced to estimate the parameters. For high-dimensional data, a linear shrinkage estimator of the covariance matrix is introduced to replace the sample covariance matrix in the likelihood. By imposing the shrinkage parameter to be common across visits and subjects, it achieves the optimal property with the uniformly minimum quadratic loss asymptotically among all linear combinations of the identity matrix and the sample covariance matrix. Asymptotic consistency of the estimators is studied for both the low-dimensional and high-dimensional scenarios. Through extensive simulation studies, the proposed approach achieves good performance in identifying the relevant components and estimating the parameters. Applying to a longitudinal resting-state fMRI dataset acquired from the ADNI study, the proposed approach identifies brain networks that demonstrate the difference between males and females at different disease stages. The findings are consistent with existing AD research and the method improves the statistical power over the analysis of cross-sectional data.

In this study, it is assumed that the covariate-related component or brain network is constant over time. To support functional dynamics that actuate behavior and cognition through changing demands, reconfiguration of brain networks occurs by compartmentalizing integrated and segregated neural processing of individual brain regions~\citep{khambhati2018beyond}. Thus, future research is to account for the variation in network composition. In the current multilevel model, the within-subject variation is fully captured by the intercept term. Another future direction is to consider the variation in covariate associated coefficients. This will enable the investigation of individual trajectories.

In many brain imaging studies, imaging scans are acquired in repeated sessions. One objective of doing so is to examine the test-retest reliability, which quantifies the stability of the repeated measurements. In fMRI studies, a commonly used measure is the intraclass correlation coefficient \cite[ICC,][]{shrout1979intraclass}, which is defined as the proportion of total variation that can be attributed to the variability between subjects. Converging evidence has shown that univariate measures derived from fMRI studies exhibit a low ICC suggesting poor test-retest reliability~\citep{noble2019decade}, while multivariate reliability is substantially greater~\citep{noble2021guide}. For example, the I2C2 showed higher reliability of functional connectivity by pooling together the variance estimates across the brain~\citep{shou2013quantifying}. In this vein, the proposed approach may offer a network-level metric of test-retest reliability, where the metric may depend on other covariates, such as the demographic variables. Another potential implementation of the proposed framework is to harmonize imaging data collected at different study sites. It has been shown that correcting for differences in covariance is also essential when analyzing fMRI data~\citep{chen2021mitigating}. The proposed framework can correct for the batch effect across sites by including sites in the regression model and treating the sites as a random factor.

\appendix
\counterwithin{figure}{section}
\counterwithin{table}{section}
\counterwithin{equation}{section}
\counterwithin{lemma}{section}
\counterwithin{theorem}{section}

\section*{Appendix}

%========================================================%
This Appendix collects the technical proof of the theorems in the main text, additional theoretical results, and additional data analysis results.
%========================================================%

%========================================================%
% Theory and Proof
%========================================================%
\section{Theory and Proof}
\label{appendix:sec:proof}

%------------------------------------------
\subsection{Proof of Theorem~\ref{thm:Sigma_solution}}
\label{appendix:sub:proof_Sigma_solution}

\begin{proof}
  For optimization problem~\eqref{eq:opt_covariance}, let
  \begin{eqnarray*}
    f &=& \frac{1}{n}\sum_{i=1}^{n}\frac{1}{V_{i}}\sum_{v=1}^{V_{i}}\mathbb{E}\left\{\bgamma^\top(\rho\mu\boldsymbol{\mathrm{I}}+(1-\rho)\bS_{iv})\bgamma-\exp(\bw_{iv}^\top\bbeta_{i})\right\}^{2} \\
    &=& \frac{1}{n}\sum_{i=1}^{n}\frac{1}{V_{i}}\sum_{v=1}^{V_{i}}\left[\rho^{2}\left\{\mu(\bgamma^\top\bgamma)-\exp(\bw_{iv}^\top\bbeta)\right\}^{2}+(1-\rho)^{2}\mathbb{E}\left\{\bgamma^\top\bS_{iv}\bgamma-\exp(\bw_{iv}^\top\bbeta_{i})\right\}^{2}\right].
  \end{eqnarray*}
  \[
    \frac{\partial f}{\partial\mu}=2\rho^{2}\frac{1}{n}\sum_{i=1}^{n}\frac{1}{V_{i}}\sum_{v=1}^{V_{i}}\left\{\mu(\bgamma^\top\bgamma)-\exp(\bw_{iv}^\top\bbeta_{i})\right\}(\bgamma^\top\bgamma)=0
  \]
  \[
    \Rightarrow \quad \mu=\frac{1}{n(\bgamma^\top\bgamma)}\sum_{i=1}^{n}\frac{1}{V_{i}}\sum_{v=1}^{V_{i}}\exp(\bw_{iv}^\top\bbeta_{i}).
  \]
  Let $\phi_{iv}^{2}=\left\{\mu(\bgamma^\top\bgamma)-\exp(\bw_{iv}^\top\bbeta_{i})\right\}^{2}$ and $\psi_{iv}^{2}=\mathbb{E}\left\{\bgamma^\top\bS_{iv}\bgamma-\exp(\bw_{iv}^\top\bbeta_{i})\right\}^{2}$,
  \[
    \frac{\partial f}{\partial \rho}=2\rho\frac{1}{n}\sum_{i=1}^{n}\frac{1}{V_{i}}\sum_{v=1}^{V_{i}}\phi_{iv}^{2}-2(1-\rho)\frac{1}{n}\sum_{i=1}^{n}\frac{1}{V_{i}}\sum_{v=1}^{V_{i}}\psi_{iv}^{2}=0
  \]
  \[
    \Rightarrow \quad \rho=\frac{\sum_{i}\sum_{v}(\psi_{vi}^{2}/nV_{i})}{\sum_{i}\sum_{v}(\phi_{iv}^{2}/nV_{i})+\sum_{i}\sum_{v}(\psi_{iv}^{2}/nV_{i})}.
  \]
  Let $\delta_{iv}^{2}=\mathbb{E}\left\{\bgamma^\top\bS_{iv}\bgamma-\mu(\bgamma^\top\bgamma)\right\}^{2}$, then $\delta_{iv}^{2}=\phi_{iv}^{2}+\psi_{iv}^{2}$.
  \vskip5pt
  Let $\phi^{2}=\sum_{i}\sum_{v}(\phi_{iv}^{2}/nV_{i})$, $\psi^{2}=\sum_{i}\sum_{v}(\psi_{iv}^{2}/nV_{i})$, and $\delta^{2}=\sum_{i}\sum_{v}(\delta_{iv}^{2}/nV_{i})$. Then, $\delta^{2}=\phi^{2}+\delta^{2}$, and
  \[
    \rho=\frac{\psi^{2}}{\phi^{2}+\psi^{2}}=\frac{\psi^{2}}{\delta^{2}}
  \]
  \[
    \Rightarrow \quad \Sigma_{iv}^{*}=\frac{\psi^{2}}{\delta^{2}}\mu\boldsymbol{\mathrm{I}}+\frac{\phi^{2}}{\delta^{2}}\bS_{iv}, \quad v=1,\dots,V_{i},~ i=1,\dots,n.
  \]
  The minimum value of $f$ is
  \begin{eqnarray*}
    && \frac{1}{n}\sum_{i=1}^{n}\frac{1}{V_{i}}\sum_{v=1}^{V_{i}}\mathbb{E}\left\{\bgamma^\top\Sigma_{iv}^{*}\bgamma-\exp(\bw_{iv}^\top\bbeta_{i})\right\}^{2} \\
    &=& \frac{1}{n}\sum_{i=1}^{n}\frac{1}{V_{i}}\sum_{v=1}^{V_{i}}\mathbb{E}\left\{\frac{\psi^{2}}{\delta^{2}}\mu\bgamma^\top\bgamma+\frac{\phi^{2}}{\delta^{2}}\bgamma^\top\bS_{iv}\bgamma-\frac{\psi^{2}+\phi^{2}}{\delta^{2}}\exp(\bw_{iv}^\top\bbeta_{i})\right\}^{2} \\
    &=& \frac{1}{n}\sum_{i=1}^{n}\frac{1}{V_{i}}\sum_{v=1}^{V_{i}}\left[\frac{\psi^{4}}{\delta^{2}}\mathbb{E}\left\{\mu\bgamma^\top\bgamma-\exp(\bw_{iv}\bbeta_{i})\right\}^{2}+\frac{\phi^{4}}{\delta^{2}}\mathbb{E}\left\{\bgamma^\top\bS_{iv}\bgamma-\exp(\bw_{iv}^\top\bbeta_{i})\right\}^{2} \right] \\
    &=& \frac{1}{n}\sum_{i=1}^{n}\frac{1}{V_{i}}\sum_{v=1}^{V_{i}}\left[\frac{\psi^{4}}{\delta^{2}}\phi_{iv}^{2}+\frac{\phi^{4}}{\delta^{2}}\psi_{iv}^{2}\right]=\frac{\psi^{4}}{\delta^{2}}\phi^{2}+\frac{\phi^{4}}{\delta^{2}}\psi^{2}=\frac{\psi^{2}\phi^{2}}{\delta^{2}}.
  \end{eqnarray*}
\end{proof}
%------------------------------------------

%------------------------------------------
\subsection{Asymptotic properties}
\label{appendix:sec:asmp_lp}

In this section, we provide more details about the asymptotic properties of the proposed estimator under the scenario of $p> \max_{i,v}T_{iv}$. This is a generalization of the discussion in \citet{zhao2021principal} to a longitudinal setting. For $v=1,\dots,V_{i}$ and $i=1,\dots,n$, it is assumed that $\Sigma_{iv}$ has the eigendecomposition $\Sigma_{iv}=\Pi_{iv}\Lambda_{iv}\Pi_{iv}^\top$, where $\Lambda_{iv}=\mathrm{diag}\{\lambda_{iv1},\dots,\lambda_{ivp}\}$ is a diagonal matrix and $\Pi_{iv}=(\bpi_{iv1},\dots,\bpi_{ivp})$ is an orthonormal rotation matrix; $\{\lambda_{iv1},\dots,\lambda_{ivp}\}$ are the eigenvalues and the columns of $\Pi_{iv}$ are the corresponding eigenvectors. Let $\bZ_{iv}=\bY_{iv}\Pi_{iv}$, where $\bY_{iv}=(\by_{iv1},\dots,\by_{ivT_{i}})^\top\in\mathbb{R}^{T_{iv}\times p}$ is the data matrix of subject $i$ at visit $v$. Under the normality assumption, the columns $\bZ_{iv}=(z_{ivtj})_{t,j}$ are uncorrelated, and the rows $\bz_{ivt}=(z_{iv1},\dots,z_{ivp})^\top\in\mathbb{R}^{p}$, for $t=1,\dots,T_{iv}$, are normally distributed with mean zero and covariance matrix $\Lambda_{iv}$.
We first present the imposed assumptions.
\begin{description}
  \item[Assumption A1] There exists a constant $C_{1}$ independent of $T_{\max}$ such that $p/T_{\max}\leq C_{1}$, where $T_{\max}=\max_{i,v}T_{iv}$.
  \item[Assumption A2] Let $N=\sum_{i=1}^{n}\sum_{v=1}^{V_{i}}T_{iv}$, $p/N\rightarrow 0$ as $n,T_{\min}\rightarrow\infty$, where $T_{\min}=\min_{i,v}T_{iv}$.
  \item[Assumption A3] There exists a constant $C_{2}$ independent of $T_{\min}$ and $T_{\max}$ such that $\sum_{j=1}^{p}\mathbb{E}(z_{iv1j}^{8})/p\leq C_{2}$, for $\forall~v\in\{1,\dots,V_{i}\}$ and $\forall~i\in\{1,\dots,n\}$.
  \item[Assumption A4] Let $\mathcal{Q}$ denote the set of all the quadruples that are made of four distinct integers between $1$ and $p$, for $\forall~v\in\{1,\dots,V_{i}\}$ and $\forall~i\in\{1,\dots,n\}$,
    \[
      \lim_{T_{iv}\rightarrow\infty}\frac{p^{2}}{T_{iv}^{2}}\frac{\sum_{(j,k,l,m)\in\mathcal{Q}}\left\{\mathrm{Cov}(z_{iv1j}z_{iv1k},z_{iv1l}z_{iv1m})\right\}^{2}}{|\mathcal{Q}|}=0,
    \]
  where $|\mathcal{Q}|$ is the cardinality of set $\mathcal{Q}$.
  \item[Assumption A5] All the covariance matrices share the same set of eigenvectors, i.e., $\Pi_{iv}=\Pi$, for $v=1,\dots,V_{i}$ and $i=1,\dots,n$. For each $\Sigma_{iv}$, there exists (at least) a column, indexed by $j_{iv}$, such that $\bgamma=\bpi_{ivj_{iv}}$ and Model~\eqref{eq:model} is satisfied.
\end{description}
Assumption A1 allows the data dimension, $p$, to be greater than the (maximum) number of observations, $T_{\max}$, and to grow at the same rate as $T_{\max}$ does. This is a common regularity condition for shrinkage estimators~\citep{ledoit2004well}. Under Assumption A2, it is guaranteed that the average sample covariance matrix, $\bar{\bS}$, is positive definite. Together with Assumption A5, the eigenvectors of $\bar{\bS}$ are consistent estimators of $\Pi$~\citep{anderson1963asymptotic}. Assumptions A3 and A4 regulate $\bz_{ivt}$ on higher-order moments, which is equivalent to imposing restrictions on the higher-order moments of $\by_{ivt}$. When the data are assumed to be normally distributed, both A3 and A4 are satisfied. Assumption A5 assumes that all the covariance matrices share the same eigenspace, though the ordering of the eigenvectors may vary.

\begin{lemma}
  For given $(\bgamma,\beta_{0i},\bbeta_{1},\beta_{0},\sigma^{2})$, as $T_{\min}\rightarrow\infty$, $\mu$, $\phi^{2}$, $\psi^{2}$, and $\delta^{2}$ are bounded.
\end{lemma}
\begin{lemma}
  For given $(\bgamma,\beta_{0i},\bbeta_{1},\beta_{0},\sigma^{2})$, as $T_{\min}\rightarrow\infty$,
    \begin{enumerate}[(i)]
      \item $\mathbb{E}(\hat{\delta}_{iv}^{2}-\delta_{iv}^{2})^{2}\rightarrow 0$, for $v=1,\dots,V_{i}$ and $i=1,\dots,n$, and thus $\mathbb{E}(\hat{\delta}^{2}-\delta^{2})^{2}\rightarrow 0$;
      \item $\mathbb{E}(\hat{\psi}_{iv}^{2}-\psi_{iv}^{2})^{2}\rightarrow 0$, for $v=1,\dots,V_{i}$ and $i=1,\dots,n$, and thus $\mathbb{E}(\hat{\psi}^{2}-\psi^{2})^{2}\rightarrow 0$;
      \item $\mathbb{E}(\hat{\phi}_{iv}^{2}-\phi_{iv}^{2})^{2}\rightarrow 0$, for $v=1,\dots,V_{i}$ and $i=1,\dots,n$, and thus $\mathbb{E}(\hat{\phi}^{2}-\phi^{2})^{2}\rightarrow 0$.
    \end{enumerate}
\end{lemma}
\begin{theorem}
  For $\forall~v\in\{1,\dots,V_{i}\}$ and $\forall~i\in\{1,\dots,n\}$, $\bS_{iv}^{*}$ is a consistent estimator of $\Sigma_{iv}^{*}$, that is, as $T_{\min}=\min_{i,v}T_{iv}\rightarrow\infty$,
    \[
      \mathbb{E}\|\bS_{iv}^{*}-\Sigma_{iv}^{*}\|^{2}\rightarrow 0.
    \]
  Thus, the asymptotic expected loss of $\bS_{iv}^{*}$ and $\Sigma_{iv}^{*}$ are identical, that is,
    \[
      \mathbb{E}\left\{\bgamma^\top\bS_{iv}^{*}\bgamma-\exp(\bw_{iv}^\top\bbeta_{i})\right\}^{2}-\mathbb{E}\left\{\bgamma^\top\Sigma_{iv}^{*}\bgamma-\exp(\bw_{iv}^\top\bbeta_{i})\right\}^{2}\rightarrow 0.
    \]
\end{theorem}
For given $(\bgamma,\beta_{0i},\bbeta_{1},\beta_{0},\sigma^{2})$, let $\Sigma_{iv}^{**}$ denote the solution to the following optimization problem,
\begin{eqnarray*}
  \underset{\rho_{1},\rho_{2}}{\text{minimize}} && \frac{1}{n}\sum_{i=1}^{n}\frac{1}{V_{i}}\sum_{v=1}^{V_{i}}\left\{\bgamma^\top\Sigma_{iv}^{**}\bgamma-\exp(\bw_{i}^\top\bbeta_{i})\right\}^{2}, \\
  \text{such that} && \Sigma_{iv}^{**}=\rho_{1}\boldsymbol{\mathrm{I}}+\rho_{2}\bS_{iv}, \quad \text{for } v=1,\dots,V_{i} \text{ and } i=1,\dots,n.
\end{eqnarray*}
\begin{theorem}
  $\bS_{iv}^{*}$ is a consistent estimator of $\Sigma_{iv}^{**}$, that is, as $T_{\min}=\min_{i,v}T_{iv}\rightarrow\infty$, for $v=1,\dots,V_{i}$ and $i=1,\dots,n$,
    \[
      \mathbb{E}\|\bS_{iv}^{*}-\Sigma_{iv}^{**}\|^{2}\rightarrow 0.
    \]
  Then, $\bS_{iv}^{*}$ has the same asymptotic expected loss as $\Sigma_{iv}^{**}$ does, that is
    \[
      \mathbb{E}\left\{\bgamma^\top\bS_{iv}^{*}\bgamma-\exp(\bw_{iv}^\top\bbeta_{i})\right\}^{2}-\mathbb{E}\left\{\bgamma^\top\Sigma_{iv}^{**}\bgamma-\exp(\bw_{iv}^\top\bbeta_{i})\right\}^{2}\rightarrow 0.
    \]
\end{theorem}
\begin{theorem}\label{appendix:thm:asmp_optimality}
  Assume $(\bgamma,\beta_{0i},\bbeta_{1},\beta_{0},\sigma^{2})$ is given. With fixed $n\in\mathbb{N}^{+}$ and $V_{1},\dots,V_{n}\in\mathbb{N}^{+}$, for any sequence of linear combinations $\{\hat{\Sigma}_{iv}\}_{i,v}$ of the identity matrix and the sample covariance matrix, where the combination coefficients are constant over $v\in\{1,\dots,V_{i}\}$ and $i\in\{1,\dots,n\}$, the estimator $\bS_{iv}^{*}$ verifies:
    \[
      \lim_{T\rightarrow\infty}\inf_{T_{iv}\geq T}\left[\frac{1}{n}\sum_{i=1}^{n}\mathbb{E}\left\{\bgamma^\top\hat{\Sigma}_{iv}\bgamma-\exp(\bw_{iv}^\top\bbeta_{i})\right\}^{2}-\frac{1}{n}\sum_{i=1}^{n}\mathbb{E}\left\{\bgamma^\top\bS_{iv}\bgamma-\exp(\bw_{iv}^\top\bbeta_{i})\right\}^{2}\right]\geq 0.
    \]
  In addition, every sequence of $\{\hat{\Sigma}_{iv}\}_{i,v}$ that performs as well as $\{\bS_{iv}^{*}\}_{i,v}$ is identical to $\{\bS_{iv}^{*}\}_{i,v}$ in the limit:
    \[
      \lim_{T\rightarrow\infty}\left[\frac{1}{n}\sum_{i=1}^{n}\mathbb{E}\left\{\bgamma^\top\hat{\Sigma}_{iv}\bgamma-\exp(\bw_{iv}^\top\bbeta_{i})\right\}^{2}-\frac{1}{n}\sum_{i=1}^{n}\mathbb{E}\left\{\bgamma^\top\bS_{iv}\bgamma-\exp(\bw_{iv}^\top\bbeta_{i})\right\}^{2}\right] = 0
    \]
    \[
      \Leftrightarrow \quad \mathbb{E}\|\hat{\Sigma}_{iv}-\bS_{iv}^{*}\|^{2}\rightarrow 0, \quad \text{for } v=1,\dots,V_{i} \text{ and } i=1,\dots,n.
    \]
\end{theorem}
With $V_{i}=1$ for $i=1,\dots,n$, the proof of the theorems are presented in \citet{zhao2021principal}. With $V_{i}$ fixed, the conclusions can be generalized and the proof of these theorems are analogously extended. In addition, $\bS_{iv}^{*}$ is also well-conditioned. For a choice of $\epsilon$ and some $\kappa\in(0,1)$, such that $p/T_{\max}\rightarrow c\leq 1-\kappa$,
\[
  \mathbb{P}\left\{\lambda_{\min}(\bS_{iv}^{*})\geq\frac{1-\kappa}{2(2C_{2}+C_{1}\sqrt{C_{2}})}\right\}\rightarrow 1,
\]
where $\lambda_{\min}(\bA)$ is the minimum eigenvalue of a matrix $\bA$.

\begin{lemma}\label{appendix:lemma:beta}
  For given $\bgamma$, assume the linear shrinkage estimator, $\Sigma_{iv}^{*}$, satisfies
    \[
      \mathbb{E}(\bgamma^\top\Sigma_{iv}^{*}\bgamma)=\exp(\bw_{iv}^\top\bbeta_{i}^{*}), \quad \text{for } v=1,\dots,V_{i} \text{ and } i=1,\dots,n,
    \]
  and thus, as $n\rightarrow\infty$,
    \[
      \frac{1}{n}\sum_{i=1}^{n}\bbeta_{i}^{*}\rightarrow\bbeta,
    \]
  where $\bbeta_{i}^{*}=(\beta_{0i}^{*},\bbeta_{1}^{*\top})^\top\in\mathbb{R}^{q+1}$ and $\bbeta=(\beta_{0},\bbeta_{1}^\top)^\top\in\mathbb{R}^{q+1}$.
\end{lemma}
\begin{theorem}
  For given $\bgamma$, assume Assumptions A1--A5 are satisfied, $(\hat{\beta}_{0},\hat{\bbeta}_{1})$ is a consistent estimator of $(\beta_{0},\bbeta_{1})$ as $n,T_{\min}\rightarrow\infty$, where $T_{\min}=\min_{i,v}T_{iv}$. In addition, as $n,V_{\min},T_{\min}\rightarrow\infty$, $\hat{\sigma}^{2}$ is a consistent estimator of $\sigma^{2}$, where $V_{\min}=\min_{i}V_{i}$.
\end{theorem}
To prove Lemma~\ref{appendix:lemma:beta}, analogous to Lemma 3.3 in \citet{zhao2021principal}, we have $\bbeta_{i}^{*}=\bbeta_{i}$ conditional on $u_{i}$ and the convergence of $\bbeta$ follows. Using the consistency in Theorem~\ref{thm:asmp_sp} and Lemma~\ref{appendix:lemma:beta}, the consistency of estimating the parameters follows.
%------------------------------------------
%========================================================%

%========================================================%
% Algorithm
%========================================================%
\section{Details of Algorithm~\ref{alg:covreg}}
\label{appendix:sec:alg_covreg}

In this section, computation details of Algorithm~\ref{alg:covreg} are provided. Assuming having the output from the $s$th step, for the $(s+1)$th step, $(\beta_{01},\dots,\beta_{0n})$ and $\bbeta_{1}$ are updated following the Newton-Raphson method. For $i=1,\dots,n$,
\[
  \beta_{0i}^{(s+1)}=\beta_{0i}^{(s)}-\frac{\partial\ell/\partial\beta_{0i}^{(s)}}{\partial^{2}\ell/\partial\beta_{0i}^{(s)2}},
\]
where
\begin{eqnarray*}
  \frac{\partial\ell}{\partial\beta_{0i}} &=& \sum_{v=1}^{V_{i}}\frac{T_{iv}}{2}\left\{1-\bgamma^\top\hat{\Sigma}_{iv}\bgamma\cdot\exp(-\beta_{0i}-\bx_{iv}^\top\bbeta_{1})\right\}+\frac{(\beta_{0i}-\beta_{0})}{\sigma^{2}}, \\
  \frac{\partial^{2}\ell}{\partial\beta_{0i}^{2}} &=& \sum_{v=1}^{V_{i}}\frac{T_{iv}}{2}\left\{\bgamma^\top\hat{\Sigma}_{iv}\bgamma\cdot\exp(-\beta_{0i}-\bx_{iv}^\top\bbeta_{1})\right\}+\frac{1}{\sigma^{2}}.
\end{eqnarray*}
\[
  \bbeta_{1}^{(s+1)}=\bbeta_{1}^{(s)}-\left(\frac{\partial^{2}\ell}{\partial\bbeta_{1}^{(s)}\partial\bbeta_{1}^{(s)\top}}\right)^{-1}\frac{\partial\ell}{\partial\bbeta_{1}^{(s)}},
\]
where
\begin{eqnarray*}
  \frac{\partial\ell}{\partial\bbeta_{1}} &=& \sum_{i=1}^{n}\sum_{v=1}^{V_{i}}\frac{T_{iv}}{2}\left\{\bx_{iv}-\bgamma^\top\hat{\Sigma}_{iv}\bgamma\cdot\exp(-\beta_{0i}-\bx_{iv}^\top\bbeta_{1})\bx_{iv}\right\}, \\
  \frac{\partial^{2}\ell}{\partial\bbeta_{1}\partial\bbeta_{1}^{\top}} &=& \sum_{i=1}^{n}\sum_{v=1}^{V_{i}}\frac{T_{iv}}{2}\left\{\bgamma^\top\hat{\Sigma}_{iv}\bgamma\cdot\exp(-\beta_{0i}-\bx_{iv}^\top\bbeta_{1})\bx_{iv}\bx_{iv}^\top\right\}.
\end{eqnarray*}
For the hyperparameters $\beta_{0}$ and $\sigma^{2}$,
\[
  \frac{\partial\ell}{\partial\beta_{0}}=\sum_{i=1}^{n}\frac{-(\beta_{0i}-\beta_{0})}{\sigma^{2}}=0 \quad \Rightarrow \quad \beta_{0}^{(s+1)}=\frac{1}{n}\sum_{i=1}^{n}\beta_{0i}^{(s+1)},
\]
\[
  \frac{\partial\ell}{\partial\sigma^{2}}=\frac{n}{2\sigma^{2}}-\sum_{i=1}^{n}\frac{(\beta_{0i}-\beta_{0})^{2}}{2\sigma^{4}}=0 \quad \Rightarrow \quad \sigma^{2(s+1)}=\frac{1}{n}\sum_{i=1}^{n}\left(\beta_{0i}^{(s+1)}-\beta_{0}^{(s+1)}\right)^{2}.
\]
For $\bgamma$, it is to optimize the following problem:
\begin{eqnarray*}
  \text{minimize} && \bgamma^\top\left\{\sum_{i=1}^{n}\sum_{v=1}^{V_{i}}\frac{T_{iv}}{2}\exp(-\beta_{0i}-\bx_{iv}^\top\bbeta_{1})\hat{\Sigma}_{iv}\right\}\bgamma, \\
  \text{such that} && \bgamma^\top\bH\bgamma=1.
\end{eqnarray*}
Replacing $\hat{\Sigma}_{iv}$ with $\bS_{iv}^{*(s+1)}$, the solution is provided in Algorithm 1 in~\citet{zhao2021covariate}.
%========================================================%

%========================================================%
\section{Additional results of the ADNI study}
\label{appendix:sec:adni}

%--------------------------------------------------------
% \subsection{Results of the longitudinal resting-state fMRI data}

\begin{figure}[!ph]
  \begin{center}
    \subfloat[C1]{\includegraphics[width=0.3\textwidth]{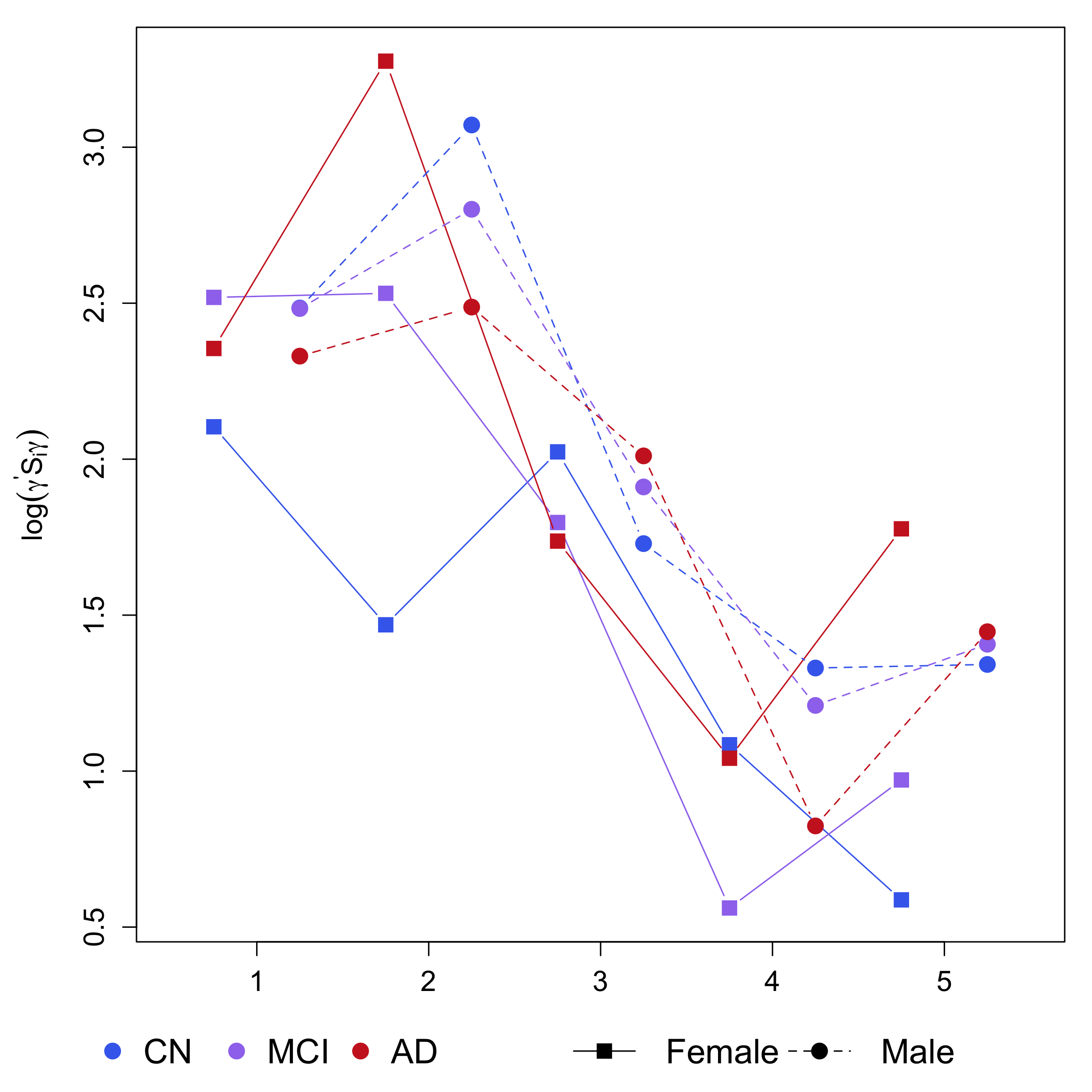}}
    \enskip{}
    \subfloat[C2]{\includegraphics[width=0.3\textwidth]{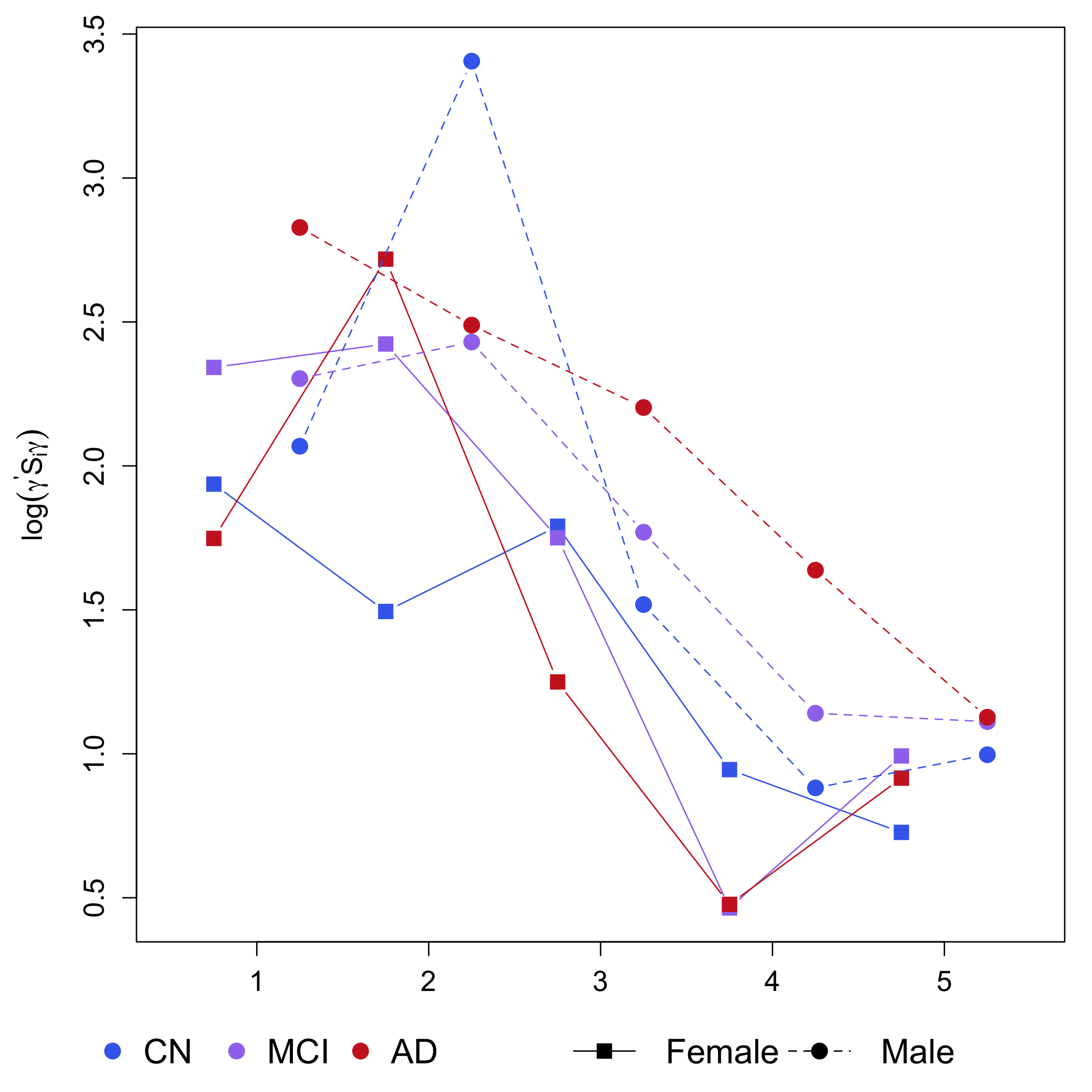}}
    \enskip{}
    \subfloat[C3]{\includegraphics[width=0.3\textwidth]{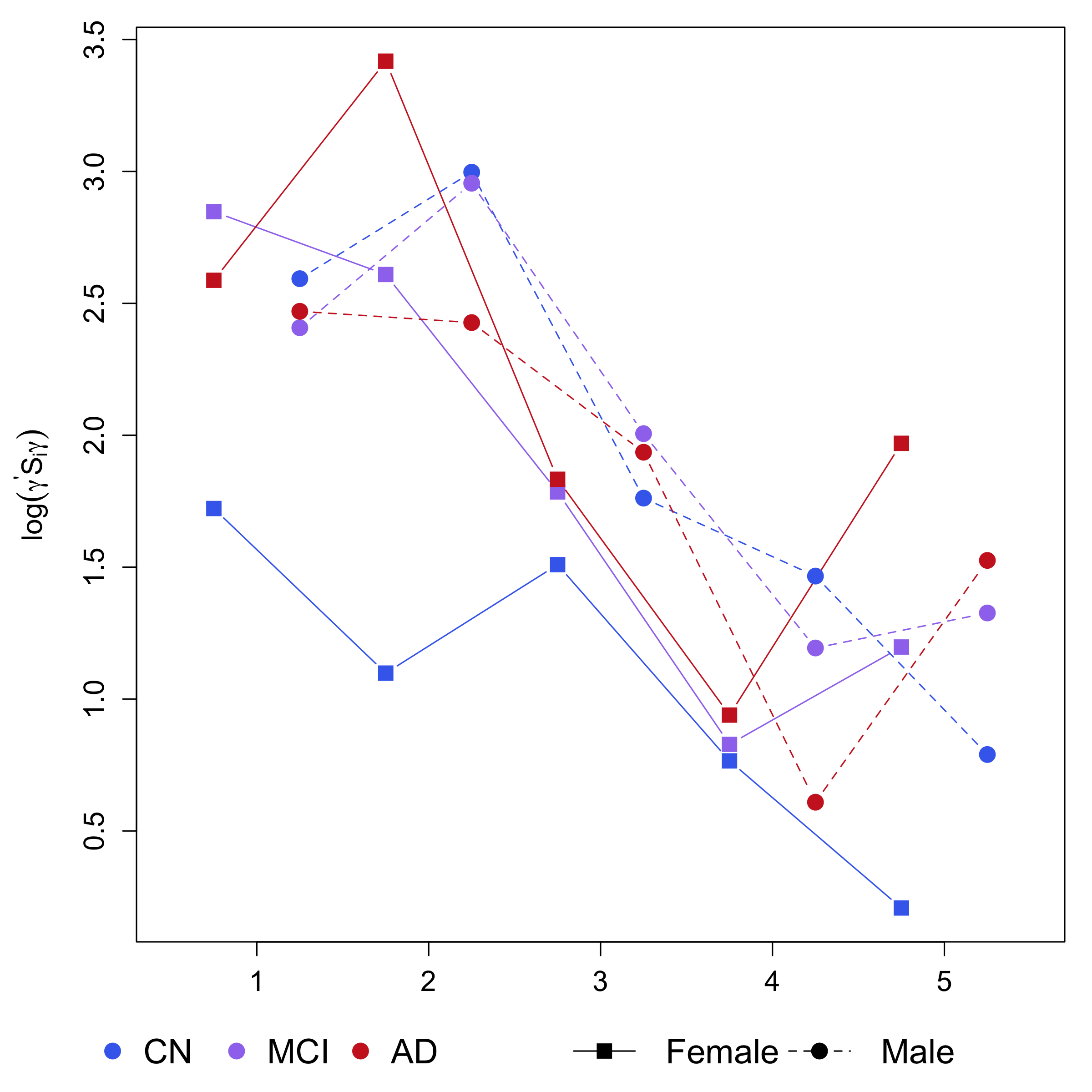}}

    \subfloat[C4]{\includegraphics[width=0.3\textwidth]{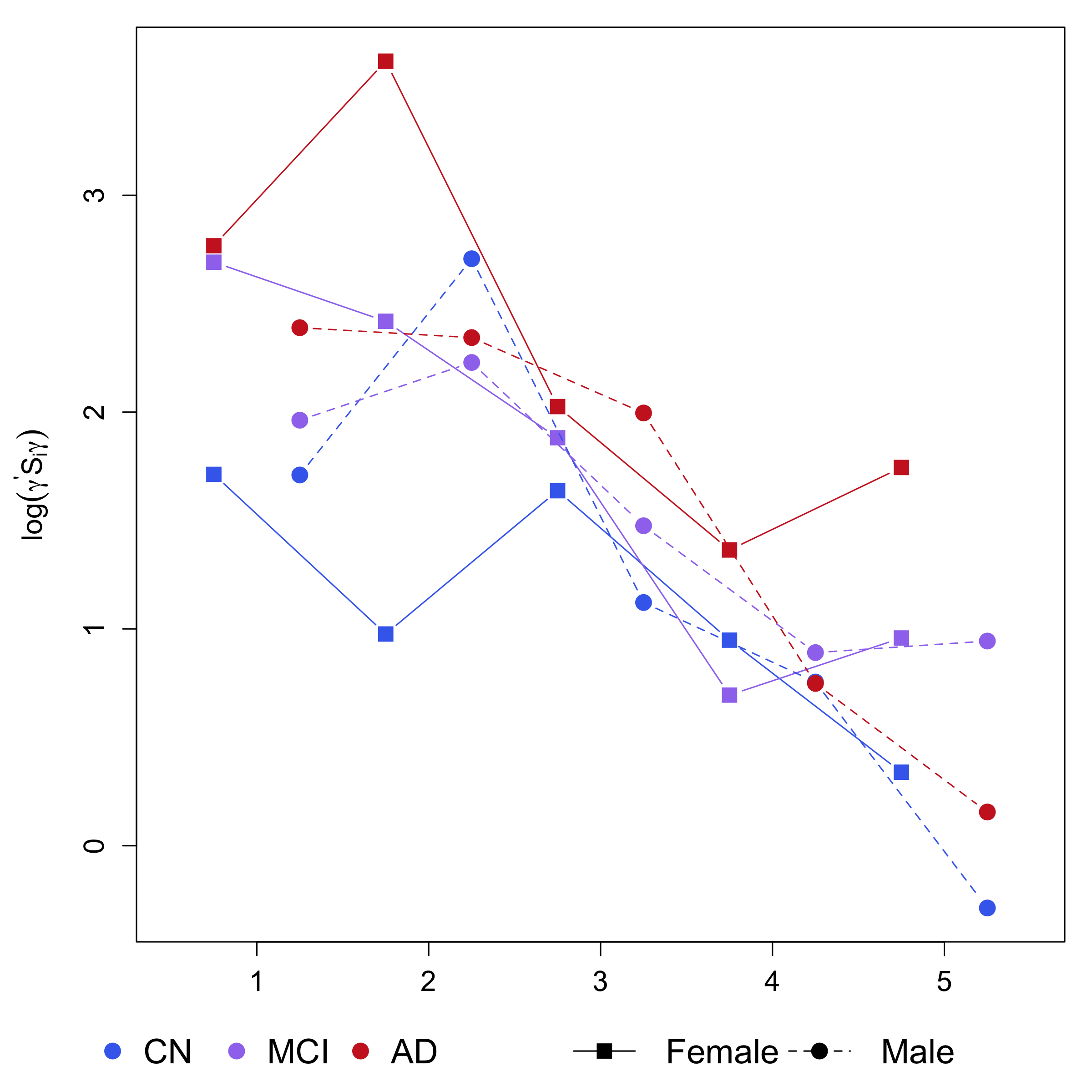}}
    \enskip{}
    \subfloat[C5]{\includegraphics[width=0.3\textwidth]{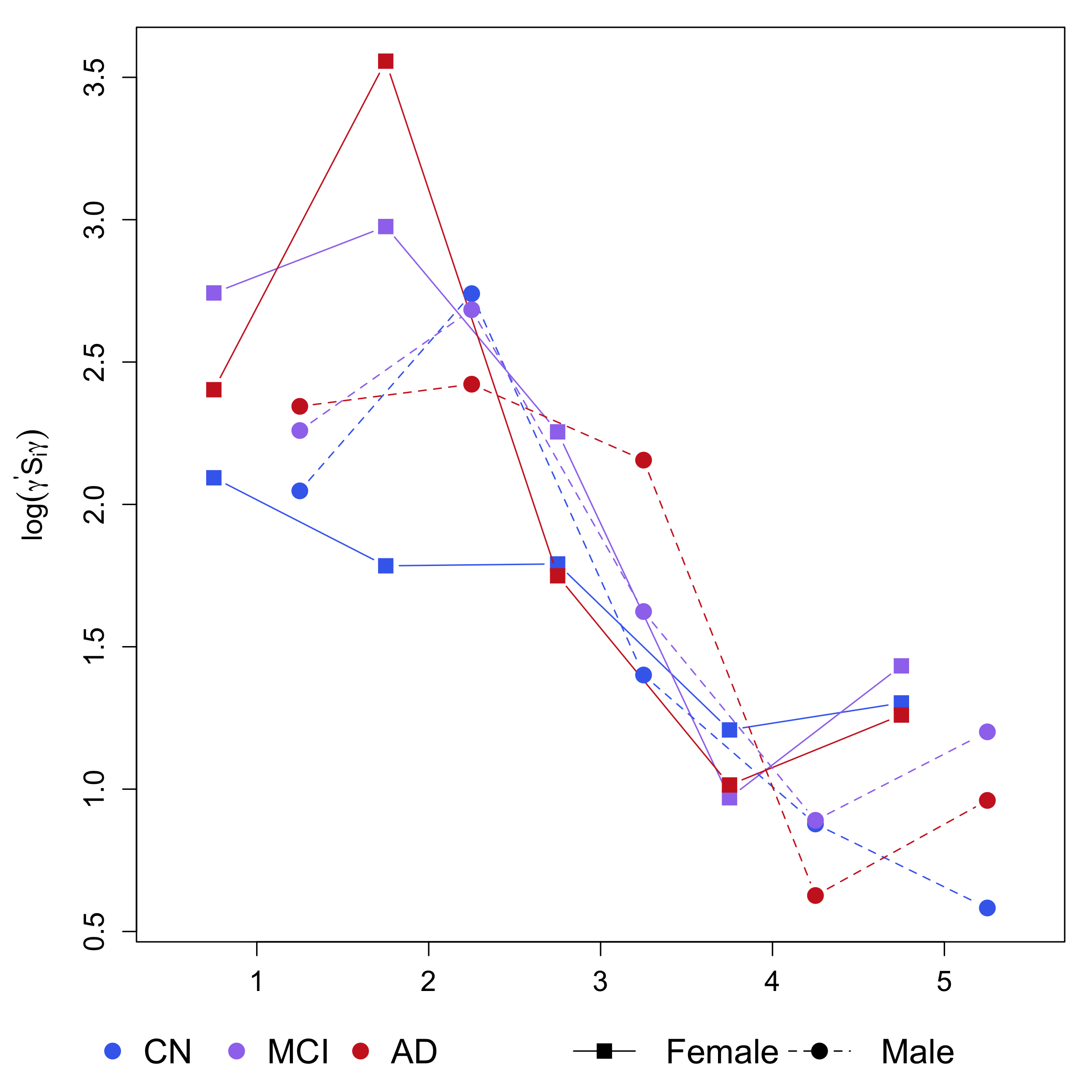}}
  \end{center}
  \caption{\label{appendix:fig:adni_scoreMean}Average $\log(\hat{\bgamma}^\top\hat{\Sigma}_{iv}\hat{\bgamma})$ outcomes of the five identified components by the diagnosis and sex subgroups at the five visits in the ADNI analysis.}
\end{figure}

\begin{figure}[!ph]
  \begin{center}
    \subfloat[C1]{\includegraphics[width=0.6\textwidth]{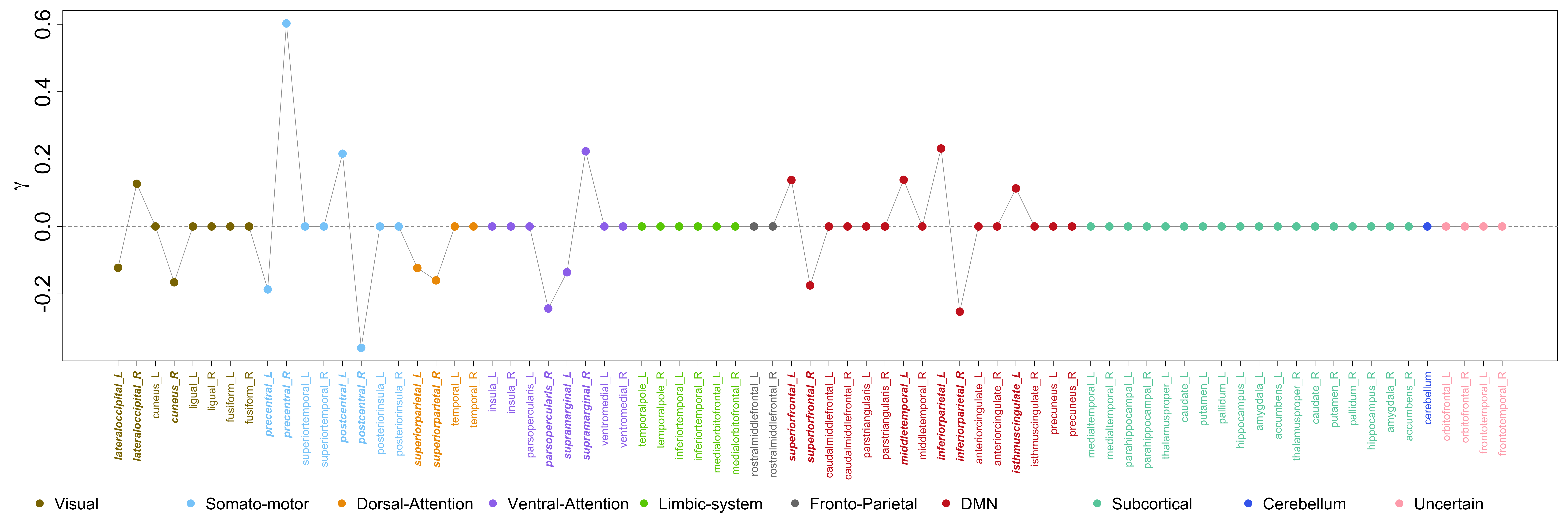}}

    \subfloat[C2]{\includegraphics[width=0.6\textwidth]{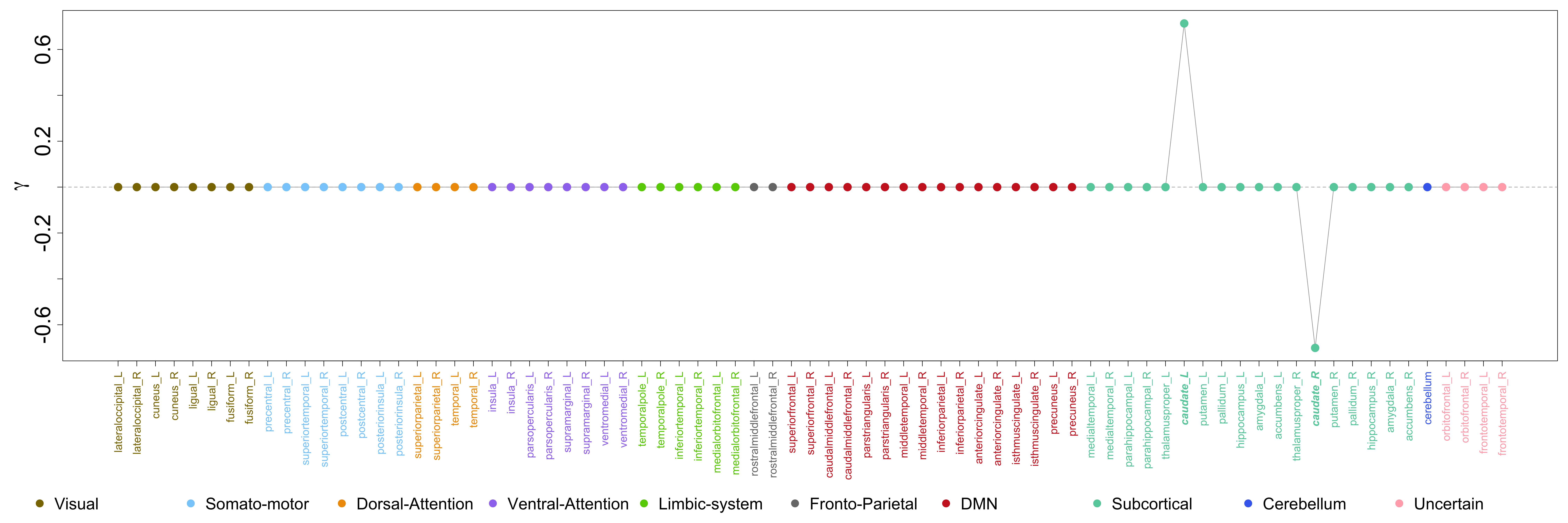}}

    \subfloat[C3]{\includegraphics[width=0.6\textwidth]{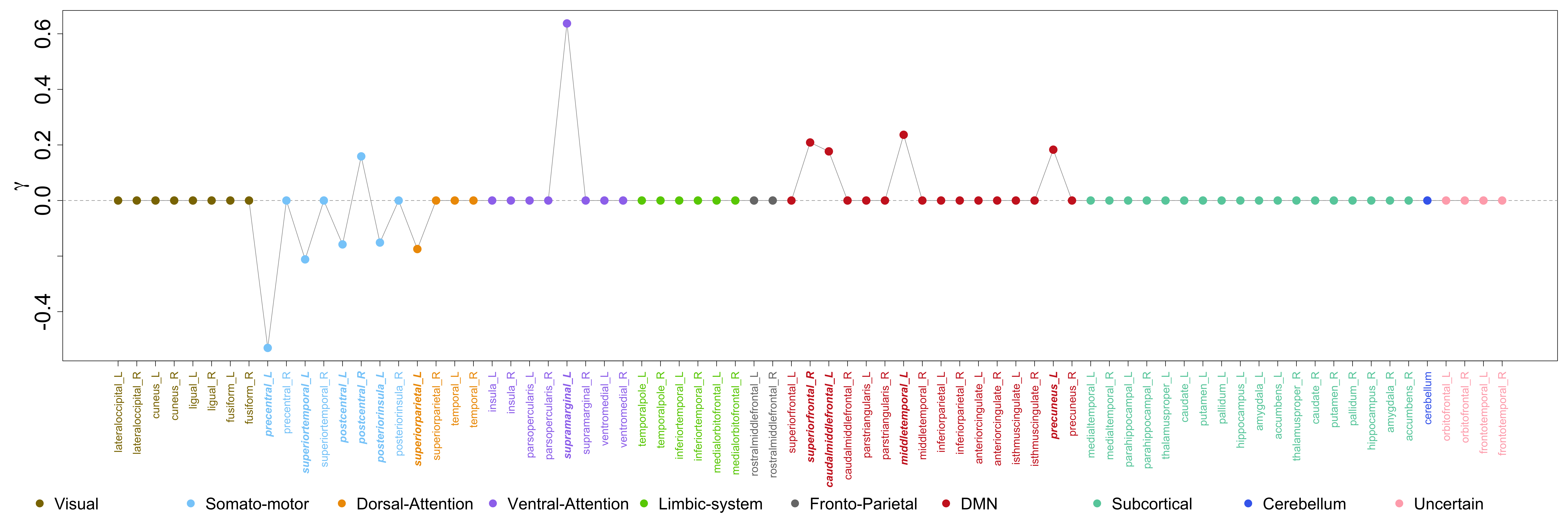}}

    \subfloat[C4]{\includegraphics[width=0.6\textwidth]{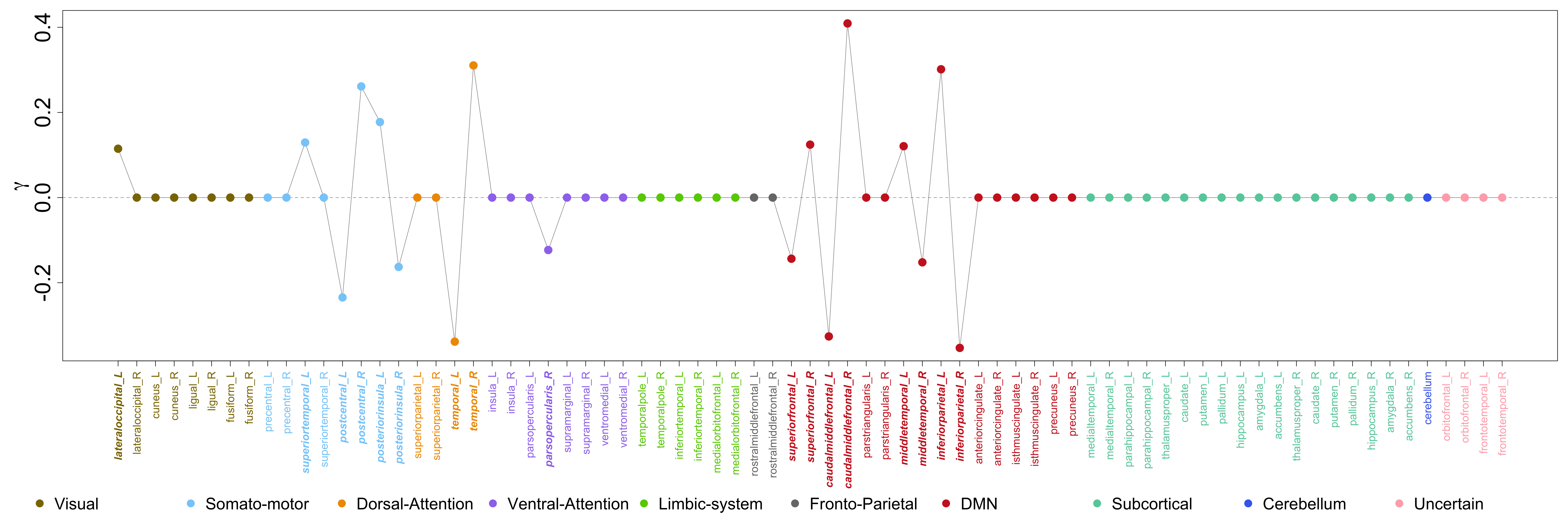}}

    \subfloat[C5]{\includegraphics[width=0.6\textwidth]{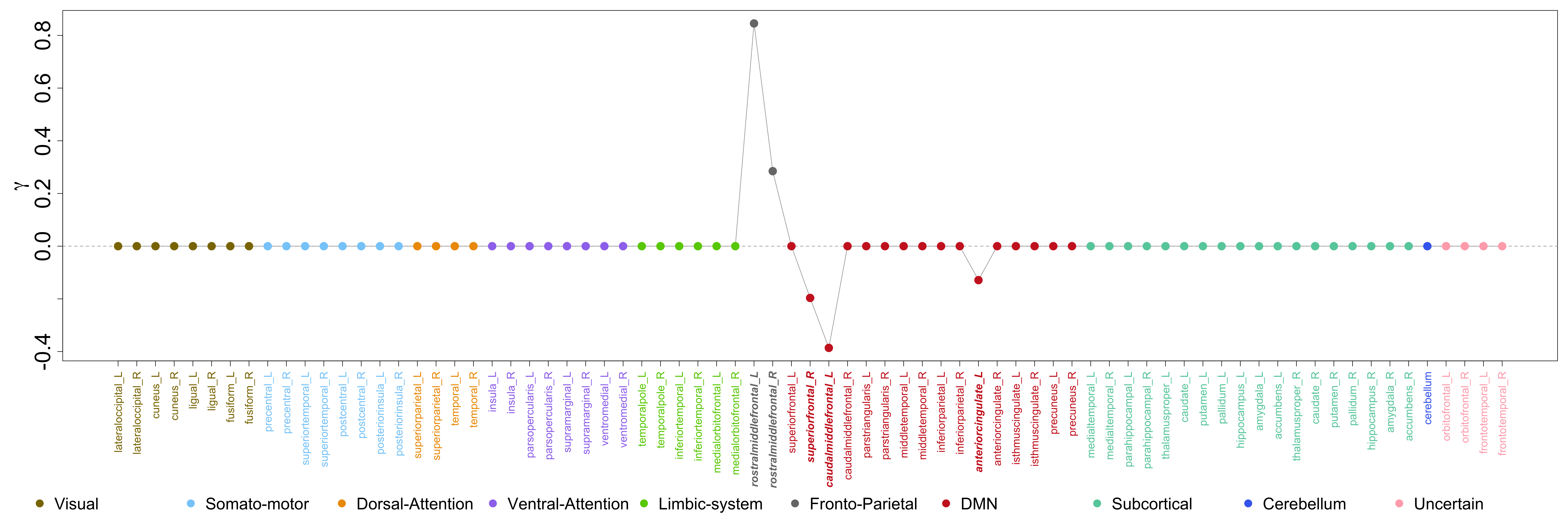}}
  \end{center}
  \caption{\label{appendix:fig:adni_loading}Sparsified loading profile of the five identified components in the ADNI analysis.}
\end{figure}

Table~\ref{appendix:table:adni_sigma2} presents the estimated within-subject variation ($\sigma^{2}$) of the five identified components in the ADNI analysis. Component C4 yields the highest variation while C1 yields the lowest.
\begin{table}[!ph]
  \caption{\label{appendix:table:adni_sigma2}The estimated within-subject variation ($\sigma^{2}$) of each identified component in the ADNI analysis.}
  \begin{center}
    \begin{tabular}{l c c c c c}
      \hline
      & \multicolumn{1}{c}{C1} & \multicolumn{1}{c}{C2} & \multicolumn{1}{c}{C3} & \multicolumn{1}{c}{C4} & \multicolumn{1}{c}{C5} \\
      \hline
      $\sigma^{2}$ & $0.273$ & $0.594$ & $0.304$ & $0.662$ & $0.442$ \\
      \hline
    \end{tabular}
  \end{center}
\end{table}
Figure~\ref{appendix:fig:adni_scoreMean} presents the longitudinal trajectory of each component's connectivity ($\log(\hat{\bgamma}^\top\hat{\Sigma}_{iv}\hat{\bgamma})$) for each diagnosis-sex subgroup over the five visits. For all components, as time progresses, the level of connectivity decreases. Subgroup differences are observed and are consistent with the results in Table~\ref{table:adni}.
Figure~\ref{appendix:fig:adni_loading} shows the sparsified loading profile of the five identified components, where a \textit{post hoc} sparsification is taken using the fused lasso penalty~\citep{tibshirani2005sparsity} by incorporating the modular information of the brain regions.

\begin{sidewaystable}
% \begin{table}
  \caption{\label{appendix:table:adni_cs}Significance of the comparisons when applying the identified components to the data of visits 1--3 (V1--V3) in the ADNI analysis. \textcolor{red!80}{\textbf{$+$}}: significant and positive; \textcolor{blue!80}{\textbf{$-$}}: significant and negative.}
  \begin{center}
    \begin{tabular}{l l c c c c c c c c c c c c c c c c c c c}
      \hline
      & & \multicolumn{3}{c}{C1} && \multicolumn{3}{c}{C2} && \multicolumn{3}{c}{C3} && \multicolumn{3}{c}{C4} && \multicolumn{3}{c}{C5} \\
      \cline{3-5}\cline{7-9}\cline{11-13}\cline{15-17}\cline{19-21}
      \multicolumn{1}{c}{\multirow{-2}{*}{Comparison}} & \multicolumn{1}{c}{\multirow{-2}{*}{Group}} & \multicolumn{1}{c}{V1} & \multicolumn{1}{c}{V2} & \multicolumn{1}{c}{V3} && \multicolumn{1}{c}{V1} & \multicolumn{1}{c}{V2} & \multicolumn{1}{c}{V3} && \multicolumn{1}{c}{V1} & \multicolumn{1}{c}{V2} & \multicolumn{1}{c}{V3} && \multicolumn{1}{c}{V1} & \multicolumn{1}{c}{V2} & \multicolumn{1}{c}{V3} && \multicolumn{1}{c}{V1} & \multicolumn{1}{c}{V2} & \multicolumn{1}{c}{V3} \\
      \hline
      & Female &  & \textcolor{red!80}{\textbf{$+$}} &  &&  & \textcolor{red!80}{\textbf{$+$}} & \textcolor{red!80}{\textbf{$+$}} && \textcolor{red!80}{\textbf{$+$}} & \textcolor{red!80}{\textbf{$+$}} & \textcolor{red!80}{\textbf{$+$}} && \textcolor{red!80}{\textbf{$+$}} & \textcolor{red!80}{\textbf{$+$}} & \textcolor{red!80}{\textbf{$+$}} && \textcolor{red!80}{\textbf{$+$}} & \textcolor{red!80}{\textbf{$+$}} & \textcolor{red!80}{\textbf{$+$}} \\
      \multirow{-2}{*}{MCI$-$CN} & Male &  &  &  &&  & \textcolor{blue!80}{\textbf{$-$}} &  &&  &  &  &&  &  &  &&  &  &  \\
      \hline
      & Female &  &  &  && \textcolor{blue!80}{\textbf{$-$}} &  &  &&  & \textcolor{red!80}{\textbf{$+$}} & \textcolor{red!80}{\textbf{$+$}} && \textcolor{red!80}{\textbf{$+$}} & \textcolor{red!80}{\textbf{$+$}} & \textcolor{red!80}{\textbf{$+$}} &&  &  &  \\
      \multirow{-2}{*}{AD$-$CN} & Male & \textcolor{blue!80}{\textbf{$-$}} &  &  &&  &  &  && \textcolor{blue!80}{\textbf{$-$}} &  &  &&  &  &  &&  &  &  \\
      \hline
      & Female &  & \textcolor{blue!80}{\textbf{$-$}} &  && \textcolor{blue!80}{\textbf{$-$}} & \textcolor{blue!80}{\textbf{$-$}} &  &&  & \textcolor{blue!80}{\textbf{$-$}} &  &&  &  &  && \textcolor{blue!80}{\textbf{$-$}} & \textcolor{blue!80}{\textbf{$-$}} & \textcolor{blue!80}{\textbf{$-$}} \\
      \multirow{-2}{*}{AD$-$MCI} & Male & \textcolor{blue!80}{\textbf{$-$}} &  &  &&  &  &  &&  &  &  &&  &  &  &&  &  &  \\
      \hline
      & CN &  &  &  &&  & \textcolor{red!80}{\textbf{$+$}} & \textcolor{red!80}{\textbf{$+$}} && \textcolor{red!80}{\textbf{$+$}} & \textcolor{red!80}{\textbf{$+$}} & \textcolor{red!80}{\textbf{$+$}} &&  & \textcolor{red!80}{\textbf{$+$}} &  &&  &  &  \\
      & MCI &  &  &  &&  &  &  && \textcolor{blue!80}{\textbf{$-$}} &  &  && \textcolor{blue!80}{\textbf{$-$}} &  &  &&  & \textcolor{blue!80}{\textbf{$-$}} & \textcolor{blue!80}{\textbf{$-$}} \\
      \multirow{-3}{*}{Male$-$Female} & AD & \textcolor{blue!80}{\textbf{$-$}} &  &  &&  &  &  && \textcolor{blue!80}{\textbf{$-$}} &  &  &&  &  &  &&  &  &  \\
      \hline
      Age & &  &  & \textcolor{blue!80}{\textbf{$-$}} && \textcolor{blue!80}{\textbf{$-$}} & \textcolor{blue!80}{\textbf{$-$}} & \textcolor{blue!80}{\textbf{$-$}} && \textcolor{blue!80}{\textbf{$-$}} & \textcolor{blue!80}{\textbf{$-$}} & \textcolor{blue!80}{\textbf{$-$}} &&  &  &  &&  &  &  \\
      \hline
    \end{tabular}
  \end{center}
% \end{table}
\end{sidewaystable}
%--------------------------------------------------------
%========================================================%
%%%%%%%%%%%%%%%%%%%%%%%%%%%%%%%%%%%%%%%%%%%%%%%%%%%%%%%%%%

%%%%%%%%%%%%%%%%%%%%%%%%%%%%%%%%%%%%%%%%%%%%%%%%%%%%%%%%%%
%========================================================%
% Reference
%========================================================%
% \clearpage

\bibliographystyle{apalike}
\bibliography{Bibliography}

\begin{thebibliography}{}

\bibitem[Andersen, 1970]{andersen1970asymptotic}
Andersen, E.~B. (1970).
\newblock Asymptotic properties of conditional maximum-likelihood estimators.
\newblock {\em Journal of the Royal Statistical Society: Series B
  (Methodological)}, 32(2):283--301.

\bibitem[Anderson, 1963]{anderson1963asymptotic}
Anderson, T.~W. (1963).
\newblock Asymptotic theory for principal component analysis.
\newblock {\em The Annals of Mathematical Statistics}, 34(1):122--148.

\bibitem[Ardekani et~al., 2016]{ardekani2016analysis}
Ardekani, B.~A., Convit, A., and Bachman, A.~H. (2016).
\newblock Analysis of the {MIRIAD} data shows sex differences in hippocampal
  atrophy progression.
\newblock {\em Journal of Alzheimer's Disease}, 50(3):847--857.

\bibitem[Badhwar et~al., 2017]{badhwar2017resting}
Badhwar, A., Tam, A., Dansereau, C., Orban, P., Hoffstaedter, F., and Bellec,
  P. (2017).
\newblock Resting-state network dysfunction in {Alzheimer's} disease: a
  systematic review and meta-analysis.
\newblock {\em Alzheimer's \& Dementia: Diagnosis, Assessment \& Disease
  Monitoring}, 8:73--85.

\bibitem[Cai et~al., 2016]{cai2016structured}
Cai, T.~T., Ren, Z., and Zhou, H.~H. (2016).
\newblock Estimating structured high-dimensional covariance and precision
  matrices: Optimal rates and adaptive estimation.
\newblock {\em Electronic Journal of Statistics}, 10(1):1--59.

\bibitem[Cavedo et~al., 2018]{cavedo2018sex}
Cavedo, E., Chiesa, P.~A., Houot, M., Ferretti, M.~T., Grothe, M.~J., Teipel,
  S.~J., Lista, S., Habert, M.-O., Potier, M.-C., Dubois, B., et~al. (2018).
\newblock Sex differences in functional and molecular neuroimaging biomarkers
  of {Alzheimer's} disease in cognitively normal older adults with subjective
  memory complaints.
\newblock {\em Alzheimer's \& Dementia}, 14(9):1204--1215.

\bibitem[Chen et~al., 2021]{chen2021mitigating}
Chen, A.~A., Beer, J.~C., Tustison, N.~J., Cook, P.~A., Shinohara, R.~T., Shou,
  H., and {The Alzheimer's Disease Neuroimaging Initiative} (2021).
\newblock Mitigating site effects in covariance for machine learning in
  neuroimaging data.
\newblock {\em Human Brain Mapping}.

\bibitem[Dai et~al., 2017]{dai2017predicting}
Dai, T., Guo, Y., and {Alzheimer's Disease Neuroimaging Initiative} (2017).
\newblock Predicting individual brain functional connectivity using a
  {Bayesian} hierarchical model.
\newblock {\em NeuroImage}, 147:772--787.

\bibitem[Davison and Hinkley, 1997]{davison1997bootstrap}
Davison, A.~C. and Hinkley, D.~V. (1997).
\newblock {\em Bootstrap methods and their application}.
\newblock Number~1. Cambridge university press.

\bibitem[Efron, 1987]{efron1987better}
Efron, B. (1987).
\newblock Better bootstrap confidence intervals.
\newblock {\em Journal of the American statistical Association},
  82(397):171--185.

\bibitem[Farahani et~al., 2019]{farahani2019application}
Farahani, F.~V., Karwowski, W., and Lighthall, N.~R. (2019).
\newblock Application of graph theory for identifying connectivity patterns in
  human brain networks: a systematic review.
\newblock {\em Frontiers in Neuroscience}, 13:585.

\bibitem[Friston, 2011]{friston2011functional}
Friston, K.~J. (2011).
\newblock Functional and effective connectivity: a review.
\newblock {\em Brain Connectivity}, 1(1):13--36.

\bibitem[Gamberger et~al., 2017]{gamberger2017identification}
Gamberger, D., Lavra{\v{c}}, N., Srivatsa, S., Tanzi, R.~E., and Doraiswamy,
  P.~M. (2017).
\newblock Identification of clusters of rapid and slow decliners among subjects
  at risk for {Alzheimer’s} disease.
\newblock {\em Scientific Reports}, 7(1):1--12.

\bibitem[Goldstein, 2011]{goldstein2011bootstrapping}
Goldstein, H. (2011).
\newblock Bootstrapping in multilevel models.
\newblock {\em Handbook of advanced multilevel analysis}, pages 163--171.

\bibitem[Holland et~al., 2013]{holland2013higher}
Holland, D., Desikan, R.~S., Dale, A.~M., and McEvoy, L.~K. (2013).
\newblock Higher rates of decline for women and apolipoprotein e $\varepsilon$4
  carriers.
\newblock {\em American Journal of Neuroradiology}, 34(12):2287--2293.

\bibitem[Hua et~al., 2010]{hua2010sex}
Hua, X., Hibar, D.~P., Lee, S., Toga, A.~W., Jack~Jr, C.~R., Weiner, M.~W.,
  Thompson, P.~M., Initiative, A. D.~N., et~al. (2010).
\newblock Sex and age differences in atrophic rates: an {ADNI} study with
  $n=1368$ {MRI} scans.
\newblock {\em Neurobiology of Aging}, 31(8):1463--1480.

\bibitem[Johnstone and Lu, 2009]{johnstone2009sparse}
Johnstone, I.~M. and Lu, A.~Y. (2009).
\newblock Sparse principal components analysis.
\newblock {\em arXiv preprint arXiv:0901.4392}.

\bibitem[Khambhati et~al., 2018]{khambhati2018beyond}
Khambhati, A.~N., Mattar, M.~G., Wymbs, N.~F., Grafton, S.~T., and Bassett,
  D.~S. (2018).
\newblock Beyond modularity: Fine-scale mechanisms and rules for brain network
  reconfiguration.
\newblock {\em NeuroImage}, 166:385--399.

\bibitem[Ledoit and Wolf, 2004]{ledoit2004well}
Ledoit, O. and Wolf, M. (2004).
\newblock A well-conditioned estimator for large-dimensional covariance
  matrices.
\newblock {\em Journal of Multivariate Analysis}, 88(2):365--411.

\bibitem[Lee and Nelder, 1996]{lee1996hierarchical}
Lee, Y. and Nelder, J.~A. (1996).
\newblock Hierarchical generalized linear models.
\newblock {\em Journal of the Royal Statistical Society. Series B
  (Methodological)}, pages 619--678.

\bibitem[Li et~al., 2021]{li2021sex}
Li, X., Zhou, S., Zhu, W., Li, X., Gao, Z., Li, M., Luo, S., Wu, X., Tian, Y.,
  and Yu, Y. (2021).
\newblock Sex difference in network topology and education correlated with sex
  difference in cognition during the disease process of {Alzheimer}.
\newblock {\em Frontiers in Aging Neuroscience}, 13:241.

\bibitem[Li et~al., 2009]{li2009lstgee}
Li, Y., Zhu, H., Chen, Y., An, H., Gilmore, J., Lin, W., and Shen, D. (2009).
\newblock {LSTGEE}: Longitudinal analysis of neuroimaging data.
\newblock In {\em Medical Imaging 2009: Image Processing}, volume 7259, page
  72590F. International Society for Optics and Photonics.

\bibitem[Lin et~al., 2015]{lin2015marked}
Lin, K.~A., Choudhury, K.~R., Rathakrishnan, B.~G., Marks, D.~M., Petrella,
  J.~R., Doraiswamy, P.~M., Initiative, A. D.~N., et~al. (2015).
\newblock Marked gender differences in progression of mild cognitive impairment
  over 8 years.
\newblock {\em Alzheimer's \& Dementia: Translational Research \& Clinical
  Interventions}, 1(2):103--110.

\bibitem[Madhyastha et~al., 2018]{madhyastha2018current}
Madhyastha, T., Peverill, M., Koh, N., McCabe, C., Flournoy, J., Mills, K.,
  King, K., Pfeifer, J., and McLaughlin, K.~A. (2018).
\newblock Current methods and limitations for longitudinal {fMRI} analysis
  across development.
\newblock {\em Developmental Cognitive Neuroscience}, 33:118--128.

\bibitem[Noble et~al., 2019]{noble2019decade}
Noble, S., Scheinost, D., and Constable, R.~T. (2019).
\newblock A decade of test-retest reliability of functional connectivity: A
  systematic review and meta-analysis.
\newblock {\em Neuroimage}, 203:116157.

\bibitem[Noble et~al., 2021]{noble2021guide}
Noble, S., Scheinost, D., and Constable, R.~T. (2021).
\newblock A guide to the measurement and interpretation of {fMRI} test-retest
  reliability.
\newblock {\em Current Opinion in Behavioral Sciences}, 40:27--32.

\bibitem[Ren et~al., 2010]{ren2010nonparametric}
Ren, S., Lai, H., Tong, W., Aminzadeh, M., Hou, X., and Lai, S. (2010).
\newblock Nonparametric bootstrapping for hierarchical data.
\newblock {\em Journal of Applied Statistics}, 37(9):1487--1498.

\bibitem[Shou et~al., 2013]{shou2013quantifying}
Shou, H., Eloyan, A., Lee, S., Zipunnikov, V., Crainiceanu, A., Nebel, M.,
  Caffo, B., Lindquist, M., and Crainiceanu, C.~M. (2013).
\newblock Quantifying the reliability of image replication studies: the image
  intraclass correlation coefficient ({I2C2}).
\newblock {\em Cognitive, Affective, \& Behavioral Neuroscience},
  13(4):714--724.

\bibitem[Shrout and Fleiss, 1979]{shrout1979intraclass}
Shrout, P.~E. and Fleiss, J.~L. (1979).
\newblock Intraclass correlations: uses in assessing rater reliability.
\newblock {\em Psychological Bulletin}, 86(2):420.

\bibitem[Skup et~al., 2011]{skup2011sex}
Skup, M., Zhu, H., Wang, Y., Giovanello, K.~S., Lin, J.-a., Shen, D., Shi, F.,
  Gao, W., Lin, W., Fan, Y., et~al. (2011).
\newblock Sex differences in grey matter atrophy patterns among {AD} and {aMCI}
  patients: results from {ADNI}.
\newblock {\em Neuroimage}, 56(3):890--906.

\bibitem[Smith et~al., 2004]{smith2004advances}
Smith, S.~M., Jenkinson, M., Woolrich, M.~W., Beckmann, C.~F., Behrens, T.~E.,
  Johansen-Berg, H., Bannister, P.~R., De~Luca, M., Drobnjak, I., Flitney,
  D.~E., et~al. (2004).
\newblock Advances in functional and structural {MR} image analysis and
  implementation as {FSL}.
\newblock {\em NeuroImage}, 23:S208--S219.

\bibitem[Telzer et~al., 2018]{telzer2018methodological}
Telzer, E.~H., McCormick, E.~M., Peters, S., Cosme, D., Pfeifer, J.~H., and van
  Duijvenvoorde, A.~C. (2018).
\newblock Methodological considerations for developmental longitudinal {fMRI}
  research.
\newblock {\em Developmental Cognitive Neuroscience}, 33:149--160.

\bibitem[Tibshirani et~al., 2005]{tibshirani2005sparsity}
Tibshirani, R., Saunders, M., Rosset, S., Zhu, J., and Knight, K. (2005).
\newblock Sparsity and smoothness via the fused lasso.
\newblock {\em Journal of the Royal Statistical Society: Series B (Statistical
  Methodology)}, 67(1):91--108.

\bibitem[Tifratene et~al., 2015]{tifratene2015progression}
Tifratene, K., Robert, P., Metelkina, A., Pradier, C., and Dartigues, J.~F.
  (2015).
\newblock Progression of mild cognitive impairment to dementia due to {AD} in
  clinical settings.
\newblock {\em Neurology}, 85(4):331--338.

\bibitem[Tomasi et~al., 2011]{tomasi2011methylphenidate}
Tomasi, D., Volkow, N.~D., Wang, G.-J., Wang, R., Telang, F., Caparelli, E.~C.,
  Wong, C., Jayne, M., and Fowler, J.~S. (2011).
\newblock Methylphenidate enhances brain activation and deactivation responses
  to visual attention and working memory tasks in healthy controls.
\newblock {\em Neuroimage}, 54(4):3101--3110.

\bibitem[Van~der Leeden et~al., 2008]{van2008resampling}
Van~der Leeden, R., Meijer, E., and Busing, F.~M. (2008).
\newblock Resampling multilevel models.
\newblock In {\em Handbook of Multilevel Analysis}, pages 401--433. Springer.

\bibitem[Wang and Guo, 2019]{wang2019hierarchical}
Wang, Y. and Guo, Y. (2019).
\newblock A hierarchical independent component analysis model for longitudinal
  neuroimaging studies.
\newblock {\em NeuroImage}, 189:380--400.

\bibitem[Zhao et~al., 2021a]{zhao2021principal}
Zhao, Y., Caffo, B.~S., and Luo, X. (2021a).
\newblock Principal regression for high dimensional covariance matrices.
\newblock {\em Electronic Journal of Statistics}, 15(2):4192--4235.

\bibitem[Zhao et~al., 2021b]{zhao2021whole}
Zhao, Y., Caffo, B.~S., Wang, B., Li, C.-S.~R., and Luo, X. (2021b).
\newblock A whole-brain modeling approach to identify individual and group
  variations in functional connectivity.
\newblock {\em Brain and Behavior}, 11(1):e01942.

\bibitem[Zhao et~al., 2021c]{zhao2021covariate}
Zhao, Y., Wang, B., Mostofsky, S.~H., Caffo, B.~S., and Luo, X. (2021c).
\newblock Covariate assisted principal regression for covariance matrix
  outcomes.
\newblock {\em Biostatistics}, 22(3):629--645.

\end{thebibliography}
%========================================================%
%%%%%%%%%%%%%%%%%%%%%%%%%%%%%%%%%%%%%%%%%%%%%%%%%%%%%%%%%%

%%%%%%%%%%%%%%%%%%%%%%%%%%%%%%%%%%%%%%%%%%%%%%%%%%%%%%%%%%
%%%%%%%%%%%%%%%%%%%%%%%%%%%%%%%%%%%%%%%%%%%%%%%%%%%%%%%%%%
\end{document}